\documentclass[aps,preprint,nofootinbib,preprintnumbers,eqsecnum,superscriptaddress]{revtex4-2}

\usepackage{color}

\newcommand{\nb}[1]{\color{blue}}

\usepackage[
	  pagebackref=false,
	  colorlinks=true,
      linkcolor=blue,
      urlcolor=blue,
      filecolor=black,
      citecolor=red,
      pdfstartview=FitV,
      pdftitle={},
        pdfauthor={Thomas Faulkner, Hong Liu, Mukund Rangamani},
        pdfsubject={},
        pdfkeywords={},
        pdfpagemode=None,
        bookmarksopen=true
      ]{hyperref}

\usepackage[normalem]{ulem}
\usepackage{amsmath}
\usepackage{enumerate}
\usepackage{amsfonts}
\usepackage{epsfig}
\usepackage{mathbbol}

\newcommand\half{{\ensuremath{\frac{1}{2}}}}
\newcommand\p{\ensuremath{\partial}}

\newcommand\field[1]{{\ensuremath{\mathbb{{#1}}}}}

\newcommand\vev[1]{{\ensuremath{\left\langle{#1}\right\rangle}}}

\newcommand{\RR}{\field{R}}

\newcommand{\bega}{\begin{gather}}

\newcommand{\be}{\begin{equation}}
\newcommand{\ee}{\end{equation}}
\newcommand{\bea}{\begin{eqnarray}}
\newcommand{\eea}{\end{eqnarray}}
\newcommand{\beg}{\begin{gather}}
\newcommand{\eeg}{\end{gather}}

\newcommand{\bi}{\begin{itemize}}
\newcommand{\ei}{\end{itemize}}
\newcommand{\ben}{\begin{enumerate}}
\newcommand{\een}{\end{enumerate}}
\newcommand{\bca}{\begin{cases}}
\newcommand{\eca}{\end{cases}}
\newcommand{\bln}{\begin{align}}
\newcommand{\eln}{\end{align}}
\newcommand{\bst}{\begin{split}}
\newcommand{\est}{\end{split}}
\def\ie{\begin{equation}\begin{aligned}}
\def\fe{\end{aligned}\end{equation}}
\newcommand{\bma}{\le(\begin{matrix}}
\newcommand{\ema}{\end{matrix}\ri)}

\newcommand\al{{\alpha}}
\def\b{{\beta}}
\newcommand\ep{\epsilon}
\newcommand\sig{\sigma}
\newcommand\Sig{\Sigma}
\newcommand\lam{\lambda}
\newcommand\Lam{\Lambda}
\newcommand\om{\omega}

\newcommand\ga{{\ensuremath{{\gamma}}}}
\newcommand\Ga{{\ensuremath{{\Gamma}}}}
\newcommand\de{{\ensuremath{{\delta}}}}
\newcommand\De{{\ensuremath{{\Delta}}}}

\newcommand\ka{\kappa}

\newcommand\nab{{\nabla}}

\newcommand\Lra{{\Longrightarrow}}
\newcommand\ov{\over}
\newcommand\ha{{\half}}

\def\le{\left}
\def\ri{\right}

\newcommand\sH{{\ensuremath{{\mathcal H}}}}

\newcommand\sL{{\ensuremath{{\mathcal L}}}}
\newcommand\sM{{\ensuremath{{\mathcal M}}}}

\newcommand\vx{{\vec x}}
\newcommand\vk{{\vec k}}

\newcommand{\heq}{{\;\, \triangleq \;\,}}

\begin{document}

\title{New horizon symmetries, hydrodynamics, and quantum chaos}

\preprint{MIT-CTP/5726}

\author{Maria Knysh}
\email{maria.knysh@vub.be}
\affiliation{Theoretische Natuurkunde, Vrije Universiteit Brussel (VUB) and The International Solvay Institutes, Pleinlaan 2, B-1050 Brussels, Belgium}
\author{Hong Liu}
\email{hong$_$liu@mit.edu}
\affiliation{Center for Theoretical Physics, 
Massachusetts
Institute of Technology,
77 Massachusetts Ave.,  Cambridge, MA 02139 }
\author{Natalia Pinzani-Fokeeva}
\email{n.pinzanifokeeva@gmail.com}
\affiliation{Dipartimento di Fisica e Astronomia,~Universit\'a di Firenze,~Via G. Sansone 1, I-50019, Sesto Fiorentino, Firenze, Italy}
\affiliation{INFN, Sezione di Firenze; Via G. Sansone 1, I-50019, Sesto Fiorentino, Firenze, Italy}
\date{\today}

\begin{abstract}

We generalize the formulation of horizon symmetries presented in previous literature  to include diffeomorphisms that can shift the location of the horizon. In the context of the AdS/CFT duality, we show that horizon symmetries can be interpreted on the boundary as emergent low-energy gauge symmetries. In particular, we identify a new class of horizon symmetries that extend the so-called shift symmetry, which was previously postulated for effective field theories of maximally chaotic systems. Additionally, we comment on the connections of horizon symmetries with bulk calculations of out-of-time-ordered correlation functions and the phenomenon of pole-skipping.

\vspace{5mm}
    
{Keywords: Asymptotic symmetries, horizon symmetries, black holes, hydrodynamics, chaos}
    
\end{abstract}

\maketitle

\newpage

\tableofcontents

\section{Introduction} 

Asymptotic symmetries are local symmetries that do not vanish at infinity and preserve the boundary conditions. They are believed to represent physical symmetries of the system. For instance, in the context of the AdS/CFT duality, asymptotic symmetries in an asymptotic AdS spacetime correspond to global symmetries of the boundary system.

For a black hole geometry, the focus is often on the physics outside the horizon. In this case, it is convenient to treat the event horizon as a ``boundary'' in an effective sense as it has been done, for example, in the so-called membrane paradigm~\cite{ThoPri86}. It is natural to extend the discussion of asymptotic symmetries to an event horizon and consider diffeomorphisms that preserve the horizon of a black hole geometry~\cite{Koga:2001vq,Hotta:2000gx,Hotta:2002mq,Donnay:2015abr,Donnay:2016ejv}, as well as their physical implications. See e.g.,~\cite{Eling:2016xlx,Eling:2016qvx,Penna:2017bdn,Donnay:2019jiz,Donnelly:2020xgu,Chandrasekaran:2020wwn,Pasterski:2020xvn,Marjieh:2021gln} for further discussions on this topic and~\cite{Carlip:1998wz,Bagchi:2012xr,Hawking:2016msc,Donnelly:2016auv,Afshar:2016wfy,Carlip:2017xne,Haco:2018ske} for applications to the black hole microstate counting. Since a horizon is not a genuine physical boundary, such horizon-preserving diffeomorphisms 
are not true asymptotic symmetries, and their precise physical interpretations remain unclear. Below we will follow the standard terminology and refer to them as ``horizon symmetries''. 

To probe physical interpretations and implications of horizon symmetries, it is helpful to consider them in a familiar context.  
In this paper, we consider horizon symmetries in an eternal black hole in AdS and seek a precise physical interpretation in terms of the boundary CFT. 
We will first generalize the formulation of horizon symmetries given in~\cite{Donnay:2015abr,Donnay:2016ejv,Chandrasekaran:2020wwn}, then show that 
they can be interpreted in the boundary theory as emergent low-energy gauge symmetries, and finally discuss their connections with hydrodynamics and maximal chaos. 

The main results of this paper can be summarized as follows: 

\ben 

\item We define  horizon symmetries  as bulk diffeomorphisms that preserve both the null vector field and its non-affine parameter on the horizon.
These include those that can move the location of the horizon, i.e. without restricting to diffeomorphisms that lie along the horizon as in \cite{Chandrasekaran:2018aop}.

For a stationary black hole in a suitable coordinate system, the horizon symmetries are
\bega
\label{eq:symm1}
\sigma^0 \rightarrow \sigma^0 +\lambda(\vec{\sigma})+\alpha(\vec{\sigma})e^{-\kappa_0 \sigma^0}+\tilde{\gamma}(\vec{\sigma})e^{\kappa_0 \sigma^0} \,,\\
\label{eq:symm2}
\sigma^i \rightarrow \sigma^i +\zeta^i(\vec{\sigma})+a\,\partial_i\tilde{\gamma}(\vec{\sigma})e^{\kappa_0 \sigma^0} \,,
\end{gather}
where $(\sigma^0,\vec{\sigma})$ are horizon coordinates, $\lambda$, $\alpha$, $\tilde{\gamma}$, and $\zeta^i$  are generic functions of the spatial coordinates $\vec{\sigma}$, $a$ is a constant that depends on the specific black hole metric, and $\kappa_0$ is the non-affinity parameter.
The novel symmetry is parameterized by $\tilde{\gamma}(\vec{\sigma})$ and is proportional to an exponentially growing term $e^{\ka_0\sigma^0}$. 

\item We show that the  horizon symmetries \eqref{eq:symm1} and \eqref{eq:symm2} in fact correspond to gauge symmetries of the  dual boundary \ hydrodynamic and maximally chaotic effective field theory.  
In particular, the symmetries parametrized by $\lambda$ and $\zeta^i$ correspond to certain gauge symmetries of any hydrodynamic effective field theory  defined in \cite{Haehl:2015foa,Crossley:2015evo,Haehl:2015uoc,Jensen:2017kzi,Glorioso:2017fpd}. The symmetries parameterized by $\alpha$ and $\tilde{\gamma}$ in \eqref{eq:symm1}, dubbed as shift symmetries,  correspond to symmetries of maximally chaotic systems  as proposed by \cite{Blake:2017ris,Blake:2021wqj}.

This claim relies on the fact that both the hydrodynamic \cite{Haehl:2015foa,Crossley:2015evo,Haehl:2015uoc,Jensen:2017kzi,Glorioso:2017fpd} and maximally chaotic effective field theories \cite{Blake:2017ris,Blake:2021wqj} are based on the definition of appropriate hydrodynamical variables. Their gravitational counterpart has been discussed in~\cite{Nickel:2010pr,Crossley:2015tka,deBoer:2015ija}
which identify the dual field theory fluid variables with the relative embedding between the boundary and the horizon. Horizon symmetry transformations, like \eqref{eq:symm1} and \eqref{eq:symm2}, then  lead directly to symmetry transformations of the hydrodynamical variables. Given that 
horizon symmetries are bulk diffeomorphisms, the corresponding symmetry transformations of the hydrodynamical variables
are then identified as gauge symmetries of the dual boundary hydrodynamic and chaotic effective field theory.

\item Two special features of maximally chaotic systems as the ones arising in the context of holography \cite{Shenker:2013pqa,Roberts:2014isa,Shenker:2014cwa} and SYK models \cite{Kitaev_talk,Polchinski:2016xgd,Maldacena:2016hyu,Kitaev:2017awl} are the exponential growth of Out-of-Time Ordered Correlators (OTOCs)  and the pole-skipping phenomenon \cite{Grozdanov:2017ajz,Blake:2017ris,Blake:2018leo}. We show that the horizon symmetries parameterized by $\tilde{\gamma}$ can be used to create shock wave geometries, analogous to the ones analyzed in \cite{Shenker:2013pqa,Roberts:2014isa,Shenker:2014cwa} that lead to the exponential growth of OTOCs in holographic systems.  Moreover, we will comment on the connection between horizon symmetries and the pole-skipping phenomenon in the energy density retarded two-point function.

\een

This paper is organized as follows. In Section \ref{sec:hor} we start by illustrating our setup and defining the generators that preserve the horizon structure. In particular, we will show how symmetries like \eqref{eq:symm1}  and \eqref{eq:symm2} arise as symmetries of black hole horizons. Then, we construct the holographic dual degrees of freedom for boundary hydrodynamics  in Section \ref{sec:hydro}.  We show that these objects inherit the horizon symmetries \eqref{eq:symm1} and \eqref{eq:symm2}, establishing the correspondence and the interpretation of the horizon symmetries as symmetries of the dual fluid dynamical variables.  In Section \ref{sec:chaos} we use the horizon symmetries to generate shock wave geometries and elaborate on their connection to  the chaotic features of OTOCs as well as the pole-skipping phenomenon.  We conclude with a summary and discussion in Section \ref{sec:conclusions}. Details of some of the  derivations are deferred to Appendices \ref{A.nA}-\ref{app:A}.

\section{Formulation of horizon symmetries} \label{sec:hor}

\subsection{Setup}

Consider a $(d+1)$-dimensional black hole spacetime ${\cal M}$ with a horizon\footnote{We focus on black hole horizons despite much of this discussion can be applied to any null surface.} denoted as ${\cal H}$.
The spacetime is described by a metric $G_{AB}$ with coordinates $x^A$. 
 $\sH$ can be specified by  a set of embedding functions $X^A (\sig^a)$ in $\sM$, where 
 $\sig^a = (\sigma^0, \sigma^i) = (\sig^0, \vec \sig)$ denote intrinsic coordinates on $\sH$. We will denote bulk quantities with capital latin indices $A,B,\dots$ and horizon quantities with $a,b,\dots$.
 
 The intrinsic metric on $\sH$ can be written as 
 \bega 
\label{eq:induced}
ds^2_{\cal H} = h_{ab} d\sig^a d \sig^b, \qquad h_{ab} \heq G_{AB}(X) \p_a X^A \p_bX^B  \ ,
\end{gather} 
where $"\heq"$ is an equality evaluated at the horizon.
By definition, $h_{ab}$ is degenerate, and has a single eigenvector $\hat \ell^a$ with zero eigenvalue, 
\be \label{eno}
h_{ab} \hat \ell^b = 0 \,.
\ee 
Through $X^A (\sig^a)$, we can push forward $\hat \ell^a$ to a spacetime vector field $\ell^A$,
\be \label{llw}
\ell^A \heq  \hat \ell^a \p_a X^A\, .
\ee
The vector $\ell^A$ will be assumed to have a smooth extension away from $\sH$, and our discussion below will not depend on its extension.
By definition $\ell^A$ is null  on $\sH$, that is
\be
G_{AB}\ell^A\ell^B\heq 0\,,\label{def1}
\ee
and it is normal to ${\cal H}$ as~\eqref{eno} implies that 
\be \label{normal}
\ell_A \p_a X^A \heq   0\,,  \quad \ell_A = G_{AB}\ell^B\,.
\ee
Moreover, $\ell^A$ can be shown to satisfy the geodesics equation
\be
\ell^B\nabla_B\ell^A \heq \kappa \ell^A\,,   \label{def3}
\ee 
with some scalar $\kappa$ (non-affine parameter).

There is no intrinsic way to normalize $\hat \ell^a$, and under 
a rescaling,
\begin{gather}
\hat{\ell}^a \to 
\hat \ell'^a =
e^{\rho} \hat{\ell}^a \,, \quad 
\ka \to 
e^{\rho} (\kappa + \sL_{\ell} \rho) \heq e^{\rho} ( \ka + \sL_{\hat \ell} \rho) \,,
\label{eq:rescaling}
\end{gather}
where $\rho$ is an arbitrary scalar on the horizon, and
$\sL_\ell$ and $\sL_{\hat \ell}$ denote  the Lie derivative along $\ell^A$ and $\hat \ell^a$ respectively. 
Equations~\eqref{eq:rescaling} define an equivalence relation: the pairs $(\hat \ell^a, \ka)$ related by such rescalings should be viewed as being equivalent. The equivalence class of the pair $(\hat \ell^a, \ka)$ under~\eqref{eq:rescaling} is referred to as the {\it horizon structure}, see, e.g., ~\cite{Chandrasekaran:2018aop}. Later, we will often use the freedom~\eqref{eq:rescaling} to fix
\be \label{yba0}
\ka = \ka_0 = {\rm const}  \ .
\ee

It is convenient to introduce another spacetime 1-form $n_A$ that is null on $\sH$, satisfying 
\be \label{rigs} 
\ell^A n_A \heq -1 , \quad G_{AB} n^A n^B \heq 0 \,, \quad n^A=G^{AB}n_B    \ .
\ee
We can pull back $n_A$ to $\sH$ to define a 1-form $\hat n_a$,
\be\label{eq:n.sigma}
\hat{n}_a \heq \partial_aX^A n_A\,, \quad \hat n_a \hat \ell^a = -1 \ .
\ee
Using~\eqref{rigs}, we can also define a projector 
\begin{equation}
\label{eq:completeness}
\Pi_A{^B} =\delta_A^B+\ell_A n^B \,,\quad \Pi_A{^B} \ell_B\heq 0\,,\quad  n^A 
\Pi_A{^B} \heq 0 \ .
\end{equation}
The quantity $\Pi_A{^B}$ is asymmetric; acting with it on a vector from the left projects to the space orthogonal to $n^A$, while acting from the right projects to 
the space orthogonal to $\ell_A$.  We can also define a symmetric, codimension-two, transverse projector 
\be 
q_A{^B} = \de_A^B + \ell_A n^B + n_A \ell^B \,, \quad q_A{^B} \ell_B \heq q_A{^B} n_B \heq 0\,, \quad 
\ell^A q_A{^B}  \heq n^A   q_A{^B}\heq 0 \,,
\ee
which acts as a projector onto the base space of the horizon ${\cal H}$. That is, if we assume the horizon to have a topology ${\cal H}={\cal R}\times {\cal Z}$, where ${\cal R}$ is the null direction, $q_A{}^{B}$ projects onto the space ${\cal Z}$.

Notice that the choice of $n_A$ is not unique. It is always possible to shift its value as long as \eqref{rigs} is satisfied as we discuss in detail in Appendix \ref{A.nA}.  Our discussion will not depend on a specific choice of $n_A$.

The most general horizon metric can be written in the form 
\be 
ds^2_{\cal H }= \gamma_{ij} (d \sig^i - a^i d \sig^0) (d \sig^j - a^j d \sig^0) \,,
\ee
where $\gamma_{ij}$ is a non-degenerate spatial metric.  
We then have 
\bega \label{elli}
\hat \ell^a = {\cal C} (1, a^i)  \quad \text{i.e.} \quad \hat  \ell^a= {\cal C} \le(\le({\p \ov \p \sig^0}\ri)^a + a^i \le({\p \ov \p \sig^i}\ri)^a \ri) , \\
\ell^A \heq {\cal C} \le({\p X^A \ov \p \sig^0} +a^i {\p X^A \ov \p \sig^i} \ri) \,,
\end{gather} 
where ${\cal C}$ is an arbitrary normalization.
Suppose we choose the horizon coordinates $\sig^a$ such that 
\be \label{yer0}
\hat \ell^a =\le( {\p \ov \p \sig^0}\ri)^a ,
\ee
then $a^i =0$, and 
\be
ds^2_{\cal H}=\gamma_{ij}d\sigma^i d\sigma^j\,.
\ee

\subsection{Review}
\label{sec:review}

In this subsection we review the formulation of horizon symmetries given in~\cite{Chandrasekaran:2018aop}. 
Consider an infinitesimal diffeomorphism on $\sH$ generated by a vector field $\xi^a (\sig^b)$ 
under which 
\be
\hat \ell^a \to \hat \ell'^a =\hat  \ell^a +  \sL_{\xi}\hat{\ell}^a =\hat \ell^a + [\xi, \hat \ell]^a  \,, 
\quad \ka \to \ka'= \ka+ \sL_\xi \ka=  \ka +\xi^a \p_a \ka   \ .
\ee
A general $\xi^a$ does not preserve the horizon structure specified by the equivalence class of the pair $(\hat \ell^a, \ka)$. The horizon symmetries of~\cite{Chandrasekaran:2018aop} are defined by those $\xi^a$ that preserve the horizon structure. That is, the corresponding  $(\hat \ell^{a'}, \ka')$ lies in the equivalence class of $(\hat\ell^a, \ka)$. This leads to the requirements  
\bega
\sL_{\xi}\hat{\ell}^a=b\hat{\ell}^a\,,\label{eq:lb}\\
\sL_\xi {\kappa}=b\kappa+\sL_{\hat{\ell}}b \,,
\label{eq:kb}
\end{gather}
for some infinitesimal scalar function $b$ on $\sH$. 

To see the implications of the conditions~\eqref{eq:lb} and \eqref{eq:kb}, it is convenient to decompose $\xi^a$  into a part that is along the null direction $\hat{\ell}^a$ and a part that is orthogonal to it, 
\be\label{decomp.chi}
\xi^a = f\hat{\ell}^a+\hat{Y}^a\,,\qquad \hat{Y}^a\hat{n}_a=0\,,
\ee
where $f$ and $\hat{Y}^a$ are functions of the horizon coordinates. These functions have been referred to as {\it supertranslations} and {\it superrotations} respectively, in analogy with the asymptotic symmetry structure and terminology of flat spacetime. 
It can be shown that, with the decomposition~\eqref{decomp.chi}, equations~\eqref{eq:lb} and \eqref{eq:kb} become the following  conditions on $f$ and $\hat{Y}^a$,
\bega
\sL_{\hat{\ell}}\hat{Y}^a\propto \hat{\ell}^a\,,\label{Constr1}\\
\sL_{\hat{\ell}}(\sL_{\hat{\ell}}+\kappa)f+\hat{Y}^a\sL_{\hat{\ell}}(\sL_{\hat{\ell}}+\kappa)\hat{n}_a +\hat{Y}^a\partial_a\kappa = 0\,,
\label{Constr2}
\end{gather}
see, e.g., \cite{Chandrasekaran:2018aop,Marjieh:2021gln}.
With~\eqref{Constr1} written as 
\be \label{defa}
\sL_{\hat{\ell}}\hat{Y}^a = a  \hat{\ell}^a  \,,\quad a=-\hat{n}_a\sL_{\hat{\ell}}\hat{Y}^a\,,
\ee
equation~\eqref{Constr2} can be written alternatively as 
\be\label{dvq}
\sL_{\hat{\ell}}(\sL_{\hat{\ell}}+\kappa)f+ (\sL_{\hat{\ell}} + \ka) a +\hat{Y}^a\partial_a\kappa = 0\, .
\ee
While in~\eqref{Constr2} there is an apparent dependence on $\hat n_a$ (and thus the choice of $n_A$), it 
can be explicitly verified that there is, in fact, none. See Appendix \ref{A.nA} for details.

To explicitly see the form of the horizon symmetry implied by \eqref{defa} and \eqref{dvq}, we can choose the normalization of $\hat \ell^a$ and the intrinsic coordinates $\sigma^a$ such that~\eqref{yba0} and~\eqref{yer0} hold. Then, with the simple choice
of $\hat n_a = (-1, 0)$, equation~\eqref{decomp.chi} has the form 
\be 
\xi^a \p_a = f \p_0  + \hat{Y}^i \p_i  \ .
\ee
In this case,~\eqref{defa} becomes 
\be
\partial_0\hat{Y}^i =0, \quad a=0, 
\label{eq:constr.Y}
\ee
and equation~\eqref{dvq} reduces to 
\bega
\partial_0^2 f+\kappa_0 \partial_0 f =0 \ .
\label{eq:constr.f}
\end{gather}
Equations~\eqref{eq:constr.Y}--\eqref{eq:constr.f} are solved by
\bega
\label{eq:Yandf}
\hat{Y}^i=\zeta^i(\vec{\sigma})\,,\\
\label{eq:Yandf2}
f = \lambda(\vec{\sigma})+\alpha(\vec{\sigma})e^{-\ka_0  \sigma^0}\,,
\end{gather}
where $\zeta^i$, $\lambda$ and $\alpha$ are arbitrary functions of the spatial horizon coordinates $\vec{\sigma}$.

The explicit form of the horizon symmetries \eqref{eq:Yandf} and \eqref{eq:Yandf2} has also been derived before in~\cite{Donnay:2015abr,Donnay:2016ejv}, where the authors considered black hole horizons expressed in Gaussian null coordinates and defined the horizon symmetries to be those bulk diffeomorphisms that preserve these coordinates. 
Other proposals for horizon symmetries include \cite{Afshar:2015wjm,Afshar:2016wfy,Chandrasekaran:2020wwn}.

\subsection{New horizon symmetries}
\label{S:new.hor}

In this subsection, we present our formulation of horizon symmetries that leads to additional symmetries beyond those given in \eqref{eq:Yandf} and \eqref{eq:Yandf2}. 
Consider an infinitesimal variation of the metric $G_{AB} \to G_{AB}' = G_{AB} + f_{AB}$,  which induces a variation of the metric on the horizon hypersurface $\sH$,
\be\label{newh}
\de h_{ab} \heq f_{AB} \p_a X^A \p_b X^B  \ .
\ee
We say that the variation $f_{AB}$ preserves the horizon structure if the following two conditions are satisfied: 

\ben 

\item The vector $\hat \ell^a$ on $\sH$  has still zero eigenvalue under the new horizon metric, i.e.,  
\be
\label{1'}
\de h_{ab} \hat \ell^b \heq f_{AB} \p_a X^A \p_b X^B \hat \ell^b  \heq f_{AB} \p_a X^A   \ell^B  \heq  0  , 
\ee
which is equivalent to 
\bega \label{1''}
f_{AB} \ell^B \heq  - c \ell_A , \quad c \heq f_{AB} \ell^B n^A \,,
\end{gather} 
for some infinitesimal function $c$ defined on $\sH$.

\item We require $\ka$ to be unchanged under the new metric, i.e., with $\nab_A'$ denoting the covariant derivative associated with $G_{AB}'$,  
\be\label{2'}
\ell^B \nab_B' \ell^A \heq \ka \ell^A \,,
\ee
which can also be written as 
\be 
\label{21}
\ell^B \de \nab_B \ell^A \heq 0, \quad \nab_B' = \nab_B + \de \nab_B \ .
\ee

\een

Now suppose the metric variation is generated by a diffeomorphism, i.e.,
\be \label{fxab}
f_{AB} = \sL_\chi G_{AB} \,,
\ee
for some infinitesimal vector field $\chi^A$.  We say that  $\chi^A$ generates a {\it horizon symmetry} if~\eqref{fxab} preserves the horizon structure as defined above. This requirement translates into the following two conditions
\bega
\label{cd1}
\sL_\chi \ell^A - G^{AB} \sL_\chi \ell_B \heq c \, \ell^A\,,\\
\label{ej1}
\ha \ell^B \ell^C  \sL_n  \sL_\chi G_{BC} + c \ka   \heq \sL_\ell c   \ .
\end{gather}
To derive equation \eqref{cd1}, we use that 
\be 
  \ell^A\sL_\chi  G_{AB} = \sL_\chi \ell_B -G_{AB}  \sL_\chi \ell^A \,.
 \ee
Combined with~\eqref{1''}, this expression  leads to 
\be\label{cd1m}
 \sL_\chi \ell_B -G_{AB}  \sL_\chi \ell^A  \heq - c \,\ell_B \,,
\ee
which readily gives \eqref{cd1}
with $c$ given by
\be\label{cd10} 
c \heq \ell^A n^B \sL_\chi G_{AB} \heq n^B \sL_\chi \ell_B - n_A \sL_\chi \ell^A\ .
\ee

The derivation of \eqref{ej1} is contained in Appendix~\ref{app:der}.
Using $\sL_{n}\sL_{\chi}-\sL_{\chi}\sL_{n}=\sL_{[n,\chi]}$, we can further rewrite the first term of~\eqref{ej1} as 
\bea
\label{ej2}
 \ell^A\ell^B\sL_n\sL_{\chi}G_{AB} 
&= & \ell^A\ell^B\sL_{\chi}\sL_{n}G_{AB}+\ell^A \ell^B\sL_{[n,\chi]}G_{AB}\nonumber\\ 
&= & \sL_{\chi}(\ell^A \ell^B\sL_{n}G_{AB})-2 \ell^B (\sL_{\chi}\ell^A)  (\sL_nG_{AB})+ \ell^A \ell^B\sL_{[n,\chi]}G_{AB}\,,
\label{ej3}
\eea
which will be convenient below.

\subsection{$\chi^A$ along the horizon} 

To see the implications of \eqref{cd1} and \eqref{ej1}, we consider first
$\chi^A$  along the horizon, i.e., 
\be \label{sepc}
\chi_A \ell^A \heq 0\,.
\ee 
 In this case, the diffeomorphisms at the horizon can be thought of as being intrinsic to the hypersurface and, as we will show below, our formulation of the horizon symmetries is equivalent to that of Sec.~\ref{sec:review}. In the next subsection, we will consider the general case with $\chi_A \ell^A \neq 0$, allowing us to derive new horizon symmetries.

For $\chi^A$ satisfying~\eqref{sepc}, we can ``pull it back'' to $\sH$ as follows
\begin{equation}\label{xidef}
\xi^a \heq  \chi^A P_A{^a}\,, \quad \chi^A \heq \partial_aX^A \xi^a\,, 
\end{equation}
where $P_A{^a}$ is the inverse of $\partial_aX^A$ in the subspace orthogonal to $\ell_A$ (recall that $\p_a X^A \ell_A \heq 0$)\footnote{We may say that the subspace spanned by $\p_a X^A$ does not contain $n^A$ while that spanned by $P_A{^a}$ does not contain $\ell_A$.}, i.e., 
it is the matrix satisfying 
\be 
n^A P_A{^a} \heq 0\,, \quad P_A{^a}\,\partial_aX^B \heq \Pi_A{^B}\, , \quad \p_a X^A P_A{^b} \heq \de_a^b \ . 
\ee

With~\eqref{sepc},~\eqref{1'} can be written as 
\bega
0 \heq \sL_\chi G_{AB} \,\p_a X^A   \ell^B  \heq \sL_\chi (h_{ab} \hat \ell^b)
- G_{AB} \ell^B \sL_\chi (\p_a X^A) - G_{AB} \p_a X^A  \sL_\chi \ell^B  \cr
\heq 0 +\ell_A \p_b X^A \p_a \xi^b  - h_{ab} \sL_\xi \hat \ell^b
\heq  - h_{ab} \sL_\xi \hat \ell^b\,,
\label{hty}
\end{gather} 
where we have used that for any vector of the form $V^A = v^a \p_a X^A$, 
\bega 
\sL_\chi V^A  = \chi^B \p_B V^A - V^B \p_B \chi^A \heq 
\xi^b  \p_b (\p_a X^A v^a)  - v^b  \p_b  (\p_a X^A \xi^a) \heq \p_a X^A \sL_\xi v^a  , \\
\sL_\chi \p_a X^A  = \chi^B \p_B \p_a X^A - \p_a X^B \p_B \chi^A \heq 
\xi^b  \p_b \p_a X^A   - \p_a  (\p_b X^A \xi^b) \heq -  \p_b X^A \p_a \xi^b  \ .
\end{gather} 
From~\eqref{hty} we conclude that~\eqref{1'} (and therefore \eqref{cd1}) is equivalent to
\be\label{cd2} 
\sL_\xi \hat \ell^a = b \hat \ell^a \,,
\ee
for some infinitesimal function $b$,
which recovers~\eqref{eq:lb}. 

We now prove the equivalence of \eqref{21} (and therefore  \eqref{ej1}) with \eqref{eq:kb}.
Multiplying $\p_a X^A$ on both sides of~\eqref{cd2} gives 
\be \label{cd3} 
\p_a X^A  \sL_\xi \hat \ell^a  \heq  \sL_\chi \ell^A \heq b  \ell^A  \quad \to \quad b \heq - n_A \sL_\chi \ell^A \,.
 \ee
Moreover, for $\chi$ given by~\eqref{xidef}, we have automatically 
\be
\label{cd21}
\sL_\chi \ell_B  \heq \tilde{b} \ell_B\quad \to \quad \tilde{b} \heq - n^A \sL_\chi \ell_A ,
\ee
which can be seen from 
\be 
\p_a X^B \sL_\chi \ell_B  = \sL_\chi (\p_a X^B \ell_B) - \ell_B \sL_\chi \p_a X^B \heq 0  \ .
\ee
From equations~\eqref{cd10},~\eqref{cd3} and~\eqref{cd21}, we find 
  \be 
c= b - \tilde b \ .
\ee

Now, consider~\eqref{ej1} for $\chi^A$ given by~\eqref{xidef}. Various terms in~\eqref{ej3} can be evaluated as follows,
\bega
 \sL_{\chi}(\ell^A \ell^B\sL_{n}G_{AB})=2 \sL_{\chi}(\ell_A\ell^B\nabla_B n^A) \heq  - 2 \sL_{\chi}(n^A\ell^B\nabla_B \ell_A)
\heq 2 \sL_{\chi}\kappa, \\ 
\ell^B (\sL_{\chi}\ell^A)  (\sL_nG_{AB}) \heq  b \ell^B \ell^A \sL_nG_{AB} \heq 2 b \ka ,  \\
\ell^A \ell^B\sL_{[n,\chi]}G_{AB} = 2 \ell_A\ell^B \nabla_B \sL_{n}\chi^A\heq
- 2 \sL_\ell  \tilde b - 2 \sL_{n}\chi^A \ell^B \nabla_B \ell_A 
\heq - 2 \sL_\ell  \tilde b + 2 \tilde b \ka, 
\end{gather} 
where in the second line we have used~\eqref{cd3} and in the third line~\eqref{cd21}. 
Using them in~\eqref{ej1}, we then find that 
\be 
\sL_\chi \ka   -   \sL_\ell  \tilde b  - b  \ka   \heq \sL_\ell c   \quad \to \quad   \sL_\chi \ka  \heq b \ka + \sL_\ell b,
 \ee
 which is~\eqref{eq:kb} given that $\kappa$ and $b$ are functions on the horizon ${\cal H}$ and therefore 
\be 
\sL_{\chi}\kappa \heq\sL_{\xi}\kappa\,,\quad \sL_{\ell}b\heq \sL_{\hat{\ell}}b\,.
\ee

\subsection{General case}

We now consider a general vector field $\chi^A$ which  we parameterize as follows
\be\label{Znonzero}
\chi^A \heq f\ell^A +Y^A + Zn^A = \tilde \chi^A + Z n^A \,,\quad Y^An_A \heq 0\,,\quad Y^A \ell_A \heq 0\,,
\ee
where $f$, $Y^A$, and  $Z$ are functions of the horizon coordinates, $Z \heq - \chi^A\ell_A$, and $\tilde \chi^A \ell_A \heq 0$ lies along $\sH$.
The horizon symmetry conditions on $\chi^A$ are~\eqref{cd1} and~\eqref{ej1}. It can be shown that (see Appendix~\ref{app:der} for detailed derivations) equation~\eqref{cd1} 
reduces to
\bega \label{ss1} 
\sL_{\ell}Z-\ka Z\heq 0\, , \\
 \label{acd3}
 \sL_\ell Y^A  -  q^{AB} \nab_B Z + Z\eta^A_\perp 
  \heq
 a  \ell^A  \,, 
 \end{gather}
 for some function $a$ which can be written as 
 \be \label{ceh1}
 a \heq  - n_A \sL_\ell Y^A \, , 
  \ee
 while equation~\eqref{ej1} reduces to
 \bega \label{fi1}
 \sL_\ell ( \sL_\ell + \ka)  f  + ( \sL_\ell + \ka) a + 
\sL_{Y}\kappa + \eta^A_\perp \nab_A Z  + Z {\mathcal{F}}  \heq 0\, , \\
{\mathcal{F}} \heq  \ha  \ell^A \ell^B \sL_n \sL_n G_{AB} - \eta^2_\perp  - (\sL_\ell -\ka)  \lam   \,,
  \end{gather} 
with
\bega
\eta_A \equiv \ell^B  (\sL_nG_{AB}) = \sL_n \ell_A - G_{AB} \sL_n \ell^B \, , 
\quad  \lam \equiv n^A \eta_A
 \heq \ell_B n^A  \nab_A n^B \,,
  \quad \eta_\perp^A = q^A{_B} \eta^B 
 \ .
  \label{ceh2}
 \end{gather} 
We can also rewrite~\eqref{fi1} (by using~\eqref{ceh1} in~\eqref{fi1}) as    
 \bega \label{fi2}
 \sL_\ell ( \sL_\ell + \ka)  f    + 
 Y^A ( \sL_\ell + \ka)   \sL_\ell n_A +
\sL_{Y}\kappa + 2 q^{AB}  \nab_A Z  \ell^C \nab_C n_B+ Z {\Tilde{\mathcal{F}}}\heq 0\, ,\\
{\Tilde{\mathcal{F}}} \heq \ha  \ell^A \ell^B \sL_n \sL_n G_{AB}   + 2  n_A   \ell^C \nab_C \eta^A_\perp
- (\sL_\ell -\ka)  \lam  \ .
  \end{gather}   

Equations \eqref{ss1}, \eqref{acd3}, and \eqref{fi1} constitute a central result of our work. They are constraint equations that select a specific class of diffeomorphisms which we call 
 horizon symmetries. They are parameterized by the supertranslations $f$, the superrotations $Y^A$, and  the radial displacement $Z$. The part dependent on $Z$ leads to the new horizon symmetry and will play a central role in characterizing the chaotic regime of a black hole as we will see explicitly later on.

The authors of \cite{Chandrasekaran:2018aop} always worked in a regime where $\chi_A\ell^A \heq -Z = 0$  which allowed them to derive the constraint equations \eqref{Constr1} and \eqref{Constr2} that lead to \eqref{eq:Yandf} and \eqref{eq:Yandf2}. Moreover, they always imposed $c=0$ by requiring $b=\tilde{b}$. This latter condition is equivalent to a constraint on terms in $\chi^A$ away from the horizon, without affecting the form of $f$ and $Y^A$. 
The authors of \cite{Donnay:2016ejv}, while allowing a general radial displacement $Z$ and deriving similar intermediate equations to ours in Gaussian null coordinates, always effectively imposed $Z=0$ by requiring independence of their horizon symmetries from the fields parameterizing  the  horizon metric (apart from $\kappa$). After additionally requiring that $\kappa=const.$, they derived equations \eqref{eq:Yandf} and \eqref{eq:Yandf2}. Their proposal of horizon symmetries also constrained some components of $\chi^A$ away from the horizon. The authors of \cite{Adami:2022ktn,Adami:2023fbm,Adami:2024rkr,Adami:2024mtu} allowed for general radial displacement Z in the context of generic null boundaries. Nevertheless, their boundary conditions did not restrict the form of $Z$ as in \eqref{ss1}.

In the following two subsections, we derive simple solutions to the constraint equations \eqref{ss1}, \eqref{acd3}, and \eqref{fi1} by using a specific parameterization of the horizon coordinates and then focusing on an explicit example.

\subsection{Simplifications of the horizon symmetry equations}

We now further simplify~\eqref{ss1}--\eqref{acd3} and~\eqref{fi1} by choosing a convenient set of horizon coordinates. Recall that
\be 
\ell_A Y^A \heq n_A Y^A \heq 0  \ .
\ee
Therefore, 
introducing  
\be 
\hat Y^a \heq Y^A P_A{^a} , \quad Y^A \heq  \hat Y^a \p_a X^A\,,
\ee
we have 
\be 
\hat Y^a \hat n_a = 0 \ .
\ee
We should view quantities $f, \hat Y^a, Z$ all as functions defined on $\sH$, i.e., they are functions of the horizon coordinates $\sig^a$. 
The quantities introduced in~\eqref{ceh2} should be evaluated on the horizon and thus are also considered as functions of $\sig^a$.

We will again choose the normalization of $\hat \ell^a$ and the intrinsic coordinates $\sigma^a$ such that~\eqref{yba0} and~\eqref{yer0} hold. 
We will choose $n^A$ such that $\hat n_a = (-1, 0)$, and thus 
$\hat Y^a = \hat{Y}^i \de_i^a$ where $\hat{Y}^i$ is a spatial vector on the horizon. 
Similarly, we have 
\bega
q^{AB} \heq \gamma^{ij} \p_i X^A \p_j X^B, \quad \eta^A_\perp \heq \hat \eta^a \p_a X^A, \quad \hat \eta^a \heq \hat{\eta}^i \de_i^a \ .
\end{gather} 
where $\hat{\eta}^i$ is also a spatial vector on the horizon such that $\hat{\eta}^a \hat{n}_a =0$.

Equation~\eqref{ss1} then becomes, 
\be \label{sol1}
\ell^A \p_A Z \heq {\p \ov \p \sig^0} Z = \ka_0  Z \quad \to \quad Z = \ga (\vec{\sigma}) e^{ \ka_0 \sig^0}  \ .
\ee
where $\gamma$ is a generic function of the spatial horizon coordinates $\vec{\sigma}$. This exponentially growing solution is the new horizon symmetry, key to our later discussion in connection to many-body quantum chaos in Sec.~\ref{sec:chaos}.

Equation~\eqref{acd3}  can be written as \bega
\p_0 (\hat{Y}^i \p_i X^A)  - \hat{Y}^i \p_i (\p_0 X^A) -  \gamma^{ij} \p_i X^A  \p_j  Z + Z \hat{\eta}^i \p_i X^A 
 = a  \p_0 X^A \\
 \to \quad (\p_0 \hat{Y}^i  - \gamma^{ij} \p_j Z + Z \hat{\eta}^i) \p_i X^A = 0, \quad a =0\,,
 \end{gather} 
 which implies that
 \be\label{sol2}
\p_0 \hat{Y}^i  - \gamma^{ij} \p_j Z + Z \hat{\eta}^i =0 \ .
 \ee
Finally, equation~\eqref{fi1} becomes
\bega \label{sol3} 
\p_0 ( \p_0 + \ka_0)  f  + \hat{\eta}^i \p_i  Z  + Z {{\mathcal{F}}} = 0\,,  \\
{{\mathcal{F}}} \heq   \ha  \ell^A \ell^B \sL_n \sL_n G_{AB} - \gamma_{ij} \hat{\eta}^i \hat{\eta}^j  - (\p_0 -\ka_0)  \lam  \ .
 \end{gather} 
It is not possible to solve equations \eqref{sol2} and \eqref{sol3} in full generality as they still depend on the details of the horizon metric $\gamma^{ij}$, on $\hat{\eta}^i$ and $\lambda$. In the next subsection, we will choose a specific example that will allow us to obtain an explicit form of the horizon symmetries.

 \subsection{An explicit example} 

Consider the boosted black brane solution in AdS${}_{d+1}$. Its metric in Eddington-Finkelstein coordinates $(r,x^{\mu})$ has the form 
\be \label{fgm2}
ds^2 = G_{AB} dx^A dx^B =  - 2 u_\mu dx^\mu dr + \chi_{\mu \nu} dx^\mu dx^\nu, 
\ee
where $u_\mu$ is a constant vector satisfying $u^\mu u_\mu =-1$, and the indices $\mu, \nu,\dots$ are raised by the Minkowski metric $\eta^{\mu \nu}$, and 
\bega \label{bbb1}
 \chi_{\mu \nu} = - F(r) u_\mu u_\nu + g(r)  \De_{\mu \nu} , \quad 
\De_{\mu \nu} = \eta_{\mu \nu} + u_\mu u_\nu 
\,, \\
F (r) = {r^2 \ov R^2} \le(1 - {r_0^d \ov r^d} \ri), \quad g(r) = {r^2 \ov R^2}  \ .
 \label{bbb2}
\end{gather}
Here,  a constant $r_0$ is the  radial location of the horizon, $R$ is the AdS scale, and $\Delta_{\mu\nu}$ is a projector in the directions that are orthogonal to $u_{\mu}$, that is $\Delta_{\mu\nu}u^{\mu}=0$. 

With the embedding function $X^A(x^{\mu}) = (r_0,x^{\mu}) $ where the horizon coordinates $x^{\mu}$ are taken to be the same as the spacetime coordinates $x^{\mu}=(x^0, \vec{x})$,   the induced metric on the horizon is 
 \be 
 ds_{\cal H}^2 = g(r_0) \De_{\mu \nu} dx^\mu dx^\nu
 = g (r_0) (\eta_{\mu \nu} + u_\mu u_\nu) dx^\mu dx^\nu\,,
 \ee
 which gives 
 \be 
 \hat \ell^\mu = {\cal C}  u^\mu \quad \to \quad  \ell^A \heq  {\cal C}  (0, u^\mu) \,, \quad \ell_A =  {\cal C}  (1, F u_\mu) \heq  {\cal C}  (1, 0)  \,,
 \ee
with an arbitrary normalization factor ${\cal C}$. We can also readily calculate  
\bega 
\ell^B \nab_B \ell^A  \heq \ell^A\left(u^{\mu}\partial_{\mu}  {\cal C}+ {    {\cal C} \ov 2} \p_r F \right) \quad \to \quad  \ka = u^{\mu}\partial_{\mu}    {\cal C}  +  {\cal C}   \hat{\ka} \,,
\end{gather} 
where 
\be
\hat \ka = {1 \ov 2}   \p_r F \bigg|_{r_0} = {2 \pi \ov \b}, 
\ee
is the non-affine parameter for the Killing vector $\le({\p \ov \p x^0}\ri)^A$, and $\b=1/T$ is the (constant) inverse temperature of the black brane. 
For the condition~\eqref{yba0} to hold, we  choose a constant ${\cal C}$ that satisfies
\be 
\label{eq:def.c}
 {\cal C} ={\ka_0 \ov \hat \ka} = {\ka_0 \ov 2 \pi} \b  \ .
\ee

The horizon coordinates $\sig^a$ satisfying the conditions~\eqref{yer0} are obtained from $x^\mu$ via the Lorentz boost that takes $u^\mu$ to $(1,\vec{0})$, 
\bega \label{tr1}
\sig^0 = - {1 \ov  {\cal C}} u_\mu x^\mu , \quad \sig^i = x^i - u^i x^0 + {u^i \ov 1+u^0} u_j x^j\, .
\end{gather}
The inverse transformation is
\bega
x^0 = -  {\cal C}   u_0 \sig^0 + u_i \sig^i, \quad x^i =   \sig^i +  {\cal C}  u^i \sig^0 + {u^i \ov 1+u^0} u_j \sig^j \,, 
\end{gather} 
which gives the embedding functions $X^A(\sigma^a)=(r_0,x^{\mu}(\sigma^a))$.
In terms of $\sig^a$, the horizon metric has the form 
\be 
ds_{\cal H}^2 = g(r_0) d \vec{\sig}^2 
\ ,
\ee
with $\gamma_{ij}=g(r_0)\delta_{ij}$. 
We also choose 
\bega 
n^A \heq - {1 \ov   {\cal C}}   \le({\p \ov \p r} \ri)^A = -  {{\cal C}^{-1}} (1, 0) , \quad n_A \heq {1 \ov   {\cal C}} \le(0,  u_\mu \ri)\, ,
\end{gather}
which leads to the projector
\bega
q^A{_B}  \heq \de^A_B + \de^A_\mu \de_B^\nu u^\mu u_\nu -  \de^A_r \de_B^r  = \de^A_\mu \de_B^\nu \De^\mu{_\nu} \, .
\end{gather}

The infinitesimal diffeomorphism vector $\chi^A$ can then be decomposed using \eqref{Znonzero} as 
\bega \label{hos0}
\chi^A \heq   {\cal C} f u^\mu \le({\p \ov \p x^\mu}\ri)^A + Y^A   - {Z \ov  {\cal C} } \le({\p \ov \p r} \ri)^A\,, \\
Y^A \heq \hat{Y}^i \Lam_i^A , \quad \Lam_i^A \heq  \de^A_\mu \lam_i{^\mu} \,,
\end{gather} 
where we have introduced 
\bega
\Lambda_i^A=\frac{\partial X^A}{\partial \sigma^i}\,,\quad\lam_i{^\mu} \equiv  {\p x^\mu \ov \p \sig^i}  = \le(u_i,\de_i^j + {u_i u^j \ov 1+u^0}  \ri), \quad 
 u_\mu \lam_i{^\mu}  =0 , \\
\eta_{\mu \nu} \lam_{i}{^\mu} \lam_{j}{^\nu} = \de_{ij}   , \quad 
\delta^{ij} \lam_{i\mu} \lam_{j \nu}  = \De_{\mu\nu}    \ ,\quad \lambda_{i\mu}=\eta_{\mu\nu}\lambda_i^{\nu}\,.
\end{gather} 
Moreover, we have 
\bega
\sL_n G_{AB} \heq -{1 \ov {\cal C}} \p_r G_{AB} , \quad 
\eta_A = \ell^B \sL_n G_{AB}  \heq - u^\mu \p_r G_{\mu A} , \quad \lam \heq\frac{1}{ {\cal C}}G^{Ar}\partial_rG_{Ar}\heq \frac{1}{\cal C}  u^\mu \p_r G_{\mu r}\, ,  \\
\quad \eta^A_\perp = q^A{_C} G^{BC} \eta_B  \heq \hat{\eta}^i \Lam_i^A , \quad 
\hat{\eta}^i = \ga^{ij} \Lam_j^B \eta_B, \quad 
 \sL_n \sL_n G_{AB} \heq {1 \ov  {{\cal C}^{2}} } \p_r^2 G_{AB} \,,
\end{gather} 
which gives 
\bega
\hat{ \eta}^i =  - {1 \ov g (r_0)}  \delta^{ij} \lam_j^\mu u^\nu \p_r G_{\mu \nu} = 0, \quad \lam = {2\pi \ov (\ka_0\b)}u^\mu \p_r G_{\mu r} = 0\,, \\
\ha \ell^A \ell^B  \sL_n \sL_n G_{AB} \heq \ha u^\mu u^\nu \p_r^2 G_{\mu \nu} = -\ha F'' (r_0) = {d (d-3)\ov 2R^2} \ .
\end{gather}

Equation~\eqref{sol1} can now be written as  
\be 
Z =  \ga (\vec{\sig}) e^{ \ka_0\sig^0} =  \ga (\vec{\sig})  e^{- {2 \pi \ov \b} u_\mu x^\mu } \,,
\ee
where $\vec{\sig}$ is given by~\eqref{tr1}. Equation~\eqref{sol2} becomes 
\bega 
\p_0\hat{Y}^i = \frac{1}{g(r_0)}\p_i Z - Z \hat{\eta}^i =\frac{1}{g(r_0)} \p_i  \ga (\vec{\sig}) e^{ \ka_0 \sig^0}\,, \\
\Lra \quad  \hat{Y}^i
= \zeta^i  (\vec{\sig}) + {1 \ov \ka_0 g(r_0) } \p_i  \ga (\vec{\sig}) e^{- {2 \pi \ov \b} u_\mu x^\mu}  \,,
\end{gather} 
and equation~\eqref{sol3} becomes 
\bega
\p_0 ( \p_0 + \ka_0)  f = - Z {\cal F}  , \quad 
{\cal F} =  -\ha F''(r_0) ,  \\
\label{eq:Fpp}
\Lra \quad f
= \lam (\vec{\sig}) +\al (\vec{\sig}) e^{{2 \pi \ov \b} u_\mu x^\mu}+{1 \ov 4 \ka_0^2 }F''(r_0) \ga (\vec{\sig})e^{- {2 \pi \ov \b} u_\mu x^\mu} \ .
\end{gather} 

A special case is the black brane with $u_\mu = (-1, 0)$, for which we have 
\bega \label{hos1}
Z =  \ga (\vx) e^{{2 \pi \ov \b} x^0} \,, \\
\label{hos3}
Y^i = \zeta^i  (\vx) + {1 \ov \ka_0 g(r_0)}  \p_i  \ga (\vx) e^{{2 \pi \ov \b} x^0}  \,, \\
f = \lam (\vx)+\al (\vx)e^{-{2 \pi \ov \b} x^0}+ {1 \ov 4 \ka_0^2 } F''(r_0) \ga (\vx)e^{{2 \pi \ov \b} x^0} \ .
\label{hos2}
\end{gather}

In the context of two-dimensional gravities, the exponentially growing and exponentially decaying behavior have also been discussed for the reparameterization of the time coordinate at the horizon in~\cite{Eling:2016qvx,Lin:2019qwu}.

\subsection{Passive perspective} \label{sec:passive}

Before concluding this section, we comment on the passive perspective of the horizon symmetries. Instead of varying the metric with the embedding fixed as we have done in Section \ref{S:new.hor} with $G_{AB}\rightarrow G_{AB}'=G_{AB}+\sL_{\chi}G_{AB}$, we can alternatively fix the metric and change the embedding of the horizon.
In the new coordinates, the embedding functions become 
\be 
X'^A (\sig) = X^A  (\sig) + \chi^A (X (\sig)) \,,
\ee
where $\chi^A$ is the vector parameterizing linearized diffeomorphisms and should be evaluated at the horizon. 

The first condition for the horizon structure to be preserved is the same as~\eqref{1'}, i.e.,
\begin{equation}
h^{\prime}_{ab}\hat{\ell}^a=0\,,
\end{equation}
with now 
\be 
h'_{ab} \heq G_{AB} (X') \p_a X'^A \p_b X'^B  \ .
\ee
With $\hat \ell^a$ unchanged, we have 
\be 
\ell'^A \heq \p_a X'^A \hat \ell^a \,,
\ee
and the second condition~\eqref{2'} becomes 
\be 
\ell'^A \nab_A \ell'^B \heq \ka \ell'^B \ .
\ee
The two descriptions are effectively equivalent as it is shown generically in Appendix \ref{app:A}.

\section{Hydrodynamical interpretation} \label{sec:hydro}

We now consider a gravity system in asymptotically AdS$_{d+1}$ spacetime with a strongly coupled conformal field theory (CFT) dual. In particular, we will consider a black hole in AdS$_{d+1}$,  dual to a 
boundary system at a finite temperature. For definiteness we take the boundary to be $\RR^{1,d-1}$ with Minkowski metric $\eta_{\mu \nu}$.\footnote{The discussion can be generalized straightforwardly to a curved boundary metric. }
In what follows, we will argue that the horizon symmetries  can be interpreted 
as emergent gauge symmetries of an effective field theory (quantum hydrodynamics) of the boundary system.

\subsection{Review} 

We start by reviewing some necessary ingredients for a theory of (quantum) hydrodynamics. In the effective  field theory approach to hydrodynamics based on an action principle developed in ~\cite{Crossley:2015evo,Glorioso:2017fpd,Liu:2018kfw} (see also~\cite{Haehl:2015foa,Haehl:2015uoc,Jensen:2017kzi}),  the dynamical variables are maps ${\cal Y}^{\mu}(\sigma^a)$ between 
the fluid spacetime labeled by the coordinates $\sig^a = (\sig^0, \sig^i)$ and the physical spacetime with coordinates 
$y^\mu = (t, y^i)$. 
The spatial coordinates $\sig^i$ in the fluid spacetime label fluid elements while $\sig^0$ can be considered as their ``internal time''. For a fixed $\sig^i$, ${\cal Y}^\mu (\sig^0, \sig^i)$ gives the spacetime trajectory of a fluid element with label $\sig^i$, parameterized by $\sig^0$. 
We denote the inverse map of ${\cal Y}^{\mu}(\sigma^a)$ as $\Sig^a (y^\mu)$, which describes the fluid elements passing through the point $\vec{y}$ at time $t$.
The dynamical variables such as the velocity field $u^{\mu}$ and temperature $T$ of the more conventional  approach  can be expressed  in terms of ${\cal Y}^\mu (\sig^a)$ as
\begin{equation}\label{yenp}
\beta^{\mu} \equiv \beta u^{\mu} =\frac{\partial {\cal Y}^{\mu}}{\partial \sigma^0} , \quad \eta_{\mu \nu}  u^\mu  u^\nu = -1 \ .
\end{equation}
 See, e.g., \cite{Liu:2018kfw} for more details and references.
 
An important element in the formulation of the effective action of hydrodynamics is that the action should be invariant under the ``gauge symmetries'' 
\be \label{hies}
\sigma^0\rightarrow \sigma^{\prime\,0}=\sigma^0+\lambda(\vec{\sigma})\,,\quad \sigma^i\rightarrow \sigma^{\prime\,i}=\sigma^i+\zeta^i(\vec{\sigma})\,,
\ee
where $\lambda$ and $\zeta^i$ are arbitrary functions of the spatial variables $\vec{\sigma}$. The symmetries \eqref{hies} can be interpreted as a reparameterization freedom of the fluid variables and of their internal time. They played a crucial role in constraining the form of the hydrodynamic effective action and the resulting correlation functions.

For a strongly coupled CFT at finite temperature $T = 1/\b$, physics at distances and time scales $L$ much greater than $\b$ is described by hydrodynamics, usually formulated in a derivative expansion with an effective expansion parameter ${\b \ov L}$. 
Nevertheless, the recent reformulation of hydrodynamics indicates that the  theory can in fact be extended to scales $L \sim O(\b)$, well beyond the usual regime of validity of hydrodynamics. Such a theory, called  {\it quantum hydrodynamics}, is nonlocal and yet may have predictive power.

In fact, in~\cite{Blake:2017ris,Blake:2021wqj}, quantum hydrodynamics was used to postulate an effective field theory for maximally chaotic systems.\footnote{The effective field theory of~\cite{Blake:2017ris,Blake:2021wqj} was constructed for systems with only energy conservation (no momentum conservation). A full effective field theory for translationally invariant maximally chaotic systems has not yet been constructed.} 
 In addition to \eqref{hies}, it was assumed that such a theory also possesses a shift symmetry of the form
\be\label{hies2}
\sigma^0\rightarrow \sigma^{\prime\, 0}=\sigma^0+{\alpha}(\vec{\sigma})e^{- \ka_0 \sigma_0}+\tilde{\gamma} (\vec{\sigma}) e^{\ka_0  \sigma_0}\,,\quad \sigma^i\rightarrow \sigma^{\prime \,i}=\sigma^i\,,
\ee
that leads to a transformation of the velocity field as follows
\bega
\label{hies3}
\beta^{0}\rightarrow \beta^{\prime\,0}=\beta\left(1-\ka_0{\alpha}(\vec{y})e^{-{\ka_0 \ov\beta} t  }+\ka_0\tilde{\gamma}(\vec{y})e^{{\ka_0 \ov\beta} t  }\right)\,,\\
\label{hies4}
\beta^i\rightarrow \beta^{\prime\,i}= \beta^i\,,
\end{gather}
where ${\alpha}$ and $\tilde{\gamma}$ are generic functions of the boundary spatial coordinates $\vec{y}$.\footnote{In writing this equation, we imposed  $\lambda$ of \cite{Blake:2017ris} to be $\lambda=\ka_0$ to match our notation. The difference is in the choice of the equilibrium configuration. As we will see shortly, we impose $\Sigma^0=\frac{1}{\beta}t$ while \cite{Blake:2017ris} have $\Sigma^0=t$. }

\subsection{Horizon symmetries as gauge symmetries of quantum hydrodynamics} 

Here we show that 
the symmetries used to formulate 
 hydrodynamics \eqref{hies}, as well as the shift symmetry \eqref{hies2}  postulated for maximally chaotic systems, can be understood on the gravity side as horizon symmetries in the context of the AdS/CFT correspondence.

For a holographic system in a local equilibrium state, hydrodynamical variables 
$\Sig^a (y^\mu)$ (or equivalently the inverse maps ${\cal Y}^{\mu}(\sigma^a)$) can be obtained from 
the relative embedding between the boundary (with coordinates $y^\mu$) and the horizon (with coordinates $\sig^a$) as developed in~\cite{Nickel:2010pr,Crossley:2015tka,deBoer:2015ija}.
More explicitly, given a bulk metric $G_{AB}$,  consider shooting a null geodesic with tangent vector $T^A (y^\mu)$ from 
the boundary to the horizon. The geodesic leaves the boundary at $y^\mu$, and reaches the horizon $\sH$ at point $\sig^a$, establishing the map $\sig^a = \Sig^a (y^\mu)$ and its inverse $y^{\mu}={\cal Y}^{\mu}(\sigma^a)$. 

The velocity field and temperature can then be obtained from the definition \eqref{yenp} in the effective field theory approach.
Using the map ${\cal Y}^\mu (\sig^a)$, we can push forward the null vector $\hat \ell^a$ defined  on the horizon  $\sH$ to a vector $\b^\mu$ on the boundary, 
\be \label{yba1}
  {\p {\cal Y}^\mu \ov \p \sig^0}={\p {\cal Y}^\mu \ov \p \sig^a}\hat{\ell}^a  = \b^\mu\,,
\ee
where we have used our choice of the horizon coordinates~\eqref{yer0}.
Below, we will see explicitly that by choosing $\ka_0$ in~\eqref{yba0} to be 
\be \label{yba2}
\ka_0 = 2 \pi
\ee 
for an equilibrium configuration, $\b$ obtained from~\eqref{yba1} coincides with the  boundary inverse temperature. Thus,  we can interpret $\b$ and $u^\mu$ defined by~\eqref{yba1}--\eqref{yba2} respectively as the local fluid inverse temperature and velocity for a general non-equilibrium case. 

Notice that the given definition of the map ${\cal Y}^{\mu}(\sigma^a)$ in gravity is not unique.
Using the Fefferman-Graham coordinates near the boundary, we can parameterize  $T^A$ in terms of a time-like vector $a^{\mu}$  as follows
\be \label{yhge}
T^A  (y^\mu)= (1, a^\mu (y^\nu)), \quad \eta_{\mu \nu} a^\mu a^\nu = -1 \ .
\ee
The choice of  $a^\mu$ corresponds to a choice of the local Lorentz frame characterizing the boundary fluid.

We now consider more explicitly how horizon symmetries act on the hydrodynamical variables $\Sig^a (y^\mu)$ or ${\cal Y}^\mu (\sig^a)$. 
For this purpose, it is convenient to write the {bulk} metric in terms of the Eddington-Finkelstein coordinates as follows
\be \label{ef00}
ds^2 = -2 a_\mu (x) e^\eta dx^\mu dr + \chi_{\mu \nu} {(r,x)} dx^\mu dx^\nu , 
\ee
in which null geodesics represented by lines of $x^{\mu}={\rm const}$ are precisely those with the tangent vector at the boundary specified by~\eqref{yhge}, see~\cite{Bhattacharyya:2008xc_2}. We can also choose the radial coordinate $r$ so that $\sH$ lies at $r =r_0 = {\rm const}$. The induced metric at the horizon can then be written as 
\be 
ds_{\cal H}^2 = \lam_{\mu \nu}  (x) dx^\mu dx^\nu, \quad \lam_{\mu \nu} = \chi_{\mu \nu} (r=r_0) \ .
\ee
With the horizon coordinates $\sig^a$ being specified by~\eqref{yba0} and~\eqref{yer0}, the horizon embedding functions $X^A (\sig^a)$ are given by $r=r_0$ and $X^\mu (\sig^a) = x^\mu (\sig^a)$. 

In~\eqref{ef00}, the radial geodesic starting from $y^\mu$ with tangent vector~\eqref{yhge} is simply given by $x^\mu (s) = y^\mu$  with $s$  a parameter along the geodesic. That is, starting from $y^{\mu}$ on the boundary, the horizon is reached at $(r_0, y^{\mu})$ and 
the map ${\cal Y}^\mu (\sig^a)$ is then given by 
\be
{\cal Y}^{\mu}(\sigma^a)=X^{\mu}(\sigma^a)=x^{\mu}(\sigma^a) \ .
\ee
 Thus the boundary local temperature and velocity field are given by 
\be 
\b^\mu ={\p X^\mu \ov \p \sig^0} =  {\p x^\mu \ov \p \sig^0} \ .
\ee

Notice  that each gravity solution $G_{AB}$ determines a solution of $\Sig^a (y^\mu)$ and ${\cal Y}^{\mu}(\sigma^a)$, which can be defined without performing any derivative expansion. Thus the effective field theory of $\Sig^a (y^\mu)$ may be considered as  ``quantum hydrodynamics" which is valid for spacetime variations of order $\b$ and goes beyond the usual effective theory arising in the context of the fluid/gravity correspondence.

Now consider making a horizon symmetry transformation  generated by some infinitesimal vector field $\chi^A$ parameterized by~\eqref{Znonzero}. 
As discussed in Sec.~\ref{sec:passive}, we can equivalently describe the transformation  in terms of a linearized shift in the embedding functions 
\be 
X^A \to X'^A (\sig^a) = X^A (\sig^a) +\chi^A (\sig^a)
\ee
with $G_{AB}$ unchanged. More explicitly, we have 
\be 
r  (\sig^a) = r_0 \to r' (\sig)  = r_0 +\chi^r (\sig^a), \quad
x^\mu (\sig^a) \to x'^\mu (\sig^a) = x^\mu (\sig^a) +\chi^\mu (\sig^a)  \ .
\ee
Since the metric does not change, the geodesic starting at $y^\mu$ now hits the horizon at 
$(r_0' (y^\mu), y^\mu)$ where 
\be
r'_0 (y^\mu) = r_0 + \chi^r (\sig^a (y^\mu))\,, 
\ee
and $\sig^a (y^\mu)$ is the inverse of $x^\mu (\sig^a)$ evaluated at $x^\mu = y^\mu$. 
The new boundary to horizon map is given by 
\be \label{ehbp}
{\cal Y}^{\prime\,\mu} (\sig^a) = x^{\prime\,\mu} (\sig^a) ={\cal  Y}^\mu (\sig^a) +  \chi^\mu (\sig^a)\, .
\ee
The new velocity field is then
\be \label{ehbp1}
\b^{\,\mu} = \p_a {\cal Y}^{\prime\,\mu} \hat \ell^a 
= \b^\mu + \partial_0 \chi^\mu  \ .
\ee

Since horizon symmetry transformations are bulk diffeomorphisms, they take a {gravity} solution to another. Moreover, the two solutions should be physically equivalent. This implies that the corresponding transformations~\eqref{ehbp}--\eqref{ehbp1} should again be viewed as ``gauge symmetries'' of the dual effective field theory of quantum hydrodynamics. In particular, such transformations could in principle change the local boundary inverse temperature and velocity, and lead to nontrivial constraints on the hydrodynamic equations. 
Below we will look at an explicit example.

\subsection{An explicit example} 

As an illustration of the above abstract discussion, we consider the boundary system on $\RR^{1,d-1}$ 
in a thermal equilibrium at inverse temperature $\b$, which is described in the bulk by the solution~\eqref{fgm2}--\eqref{bbb2} with $u_\mu = (-1, \vec 0)$,  i.e. with $x^0 =v$, 
\bega \label{fgm20}
ds^2 = G_{AB} dx^A dx^B =  2 dv dr 
- F (r)dv^2 + g (r) d \vx^2,  \\
F (r) = {r^2 \ov R^2} \le(1 - {r_0^d \ov r^d} \ri), \quad g(r) = {r^2 \ov R^2}  \ .
\end{gather}
The horizon $\sH$ is located at the constant value $r=r_0$ and the induced metric on $\sH$ is given by 
\be 
ds^2_{\cal H} = g (r_0) d \vx^2  \ .
\ee
The boundary inverse temperature is given by 
\be \label{eq:b.bry}
\b = {4 \pi \ov F'(r_0)} = {2 \pi \ov \hat \ka} \ .
\ee

Using the already derived relations \eqref{tr1}, 
the horizon coordinates $\sig^a$ satisfying the conditions~\eqref{yba0}, \eqref{yer0}\, \eqref{eq:def.c}, and~\eqref{yba2}  are
\be 
\sig^0 = {1 \ov \b} v, \quad \sig^i = x^i  \ .
\ee
Given the metric~\eqref{fgm20}, we choose  $a^\mu = (1, \vec 0)$ for the radial null geodesics defined in~\eqref{yhge}. In this way, a point $y^{\mu}$ on the boundary is simply mapped to $x^\mu (s) = y^\mu$. We then have 
\be \label{eq01}
\Sig^0 (y^\mu) ={1 \ov \b} t, \quad \Sig^i = y^i, \quad
{\cal Y}^0 = \b \sig^0, \quad {\cal Y}^i = \sig^i, 
\ee
and thus the velocity field is given by
\be \label{eq02}
\b^\mu = \b (1, \vec 0)
\ee
with the inverse temperature $\beta$ given by  \eqref{eq:b.bry}
as anticipated around \eqref{yba2}. 

The  set of horizon symmetry transformations parameterized by the vector field  $\chi^A$ was given in~\eqref{hos0} and~\eqref{hos1}--\eqref{hos2}, which we copy here for convenience 
\bega
\chi^A = f \b \le({\p \ov \p v}\ri)^A + \hat{Y}^i  \le({\p \ov \p x^i} \ri)^A  - {Z \ov \b} \le({\p \ov \p r} \ri)^A, \\
\label{eq:transf.Z.Y}
Z =  \ga (\vec{\sig}) e^{2 \pi \sig^0} \,,\\ 
\hat{Y}^i = \zeta^i  (\vec{\sig}) + {1 \ov 2 \pi g(r_0)}  \p_i  \ga (\vec{\sig}) e^{2 \pi \sig^0}  , \\
\label{eq:transf.f}
f = \lam (\vec{\sig})+\al (\vec{\sig})e^{-2 \pi \sig^0}+ {1 \ov 16 \pi^2 } F''(r_0) \ga (\vec{\sig})e^{2 \pi \sig^0} \ .
\end{gather} 
From~\eqref{ehbp1}, we then find the transformation of the boundary inverse temperature and velocity field are given by 
\bega \label{velo1}
\b^{'\,0} (y^\mu) = \b \le(1- 2 \pi \al (\vec{y})e^{-{2 \pi\ov \b} t}+ {1 \ov 8 \pi } F''(r_0) \ga (\vec{y})e^{{2 \pi\ov \b} t}  \ri) , \\
\b^{'\,i} (y^\mu)= {1 \ov g(r_0)}  \p_i  \ga (\vec{y}) e^{{2 \pi\ov \b} t}\,.
\label{velo2}
\end{gather}

We note that $\zeta^i (\vec{\sig})$ and $\lam (\vec{\sig})$ do not lead to any change in $\b^\mu$.  They lead to a diffeomorphism on the horizon  of the form
 \be\label{eq:sigma.new.fluid}
\sigma^0\rightarrow \sigma^{\prime\,0}=\sigma^0+ \lambda(\vec{\sigma})\,,\qquad \sigma^{\prime\,i}=\sigma^ i \rightarrow \sigma^i+\zeta^i (\vec{\sigma})\,,
 \ee
which matches the field theory symmetry of hydrodynamics \eqref{hies}. The velocity field transformation \eqref{velo1} matches the previous postulate \eqref{hies3} with our choice of $\ka_0$ given in \eqref{yba2} and $\tilde{\gamma}(\vec{\sig})={1 \ov 16 \pi^2 } F''(r_0) \ga (\vec{\sig})$. Equation \eqref{velo2} is instead a new prediction with respect to \eqref{hies4}.

\section{Implications for quantum chaos} \label{sec:chaos}

In Section \ref{sec:hydro}, we discussed how horizon symmetries formulated in Section~\ref{sec:hor}  lead to low-energy gauge symmetries in the effective field theory (quantum hydrodynamics) of the boundary theory {in the context of the AdS/CFT correspondence}. The symmetry transformations associated with the parameters $\lambda(\vec{\sigma})$ and $\zeta^i(\vec{\sigma})$ are related to symmetries of the dual hydrodynamic theory \cite{Glorioso:2017fpd}.
The symmetry transformations  associated with parameters 
$\al (\vec{\sig})$ and $\ga (\vec{\sig})$ are related to the shift symmetries \eqref{hies2} postulated in~\cite{Blake:2017ris,Blake:2021wqj} for maximally chaotic quantum many-body systems as we have seen explicitly towards the end of the previous section.
We now comment in more detail on the implications of these symmetries on quantum chaos by considering two probes, the OTOCs and the phenomenon of pole-skipping.

In the effective field theory formulation of  \cite{Blake:2017ris,Blake:2021wqj}, the shift symmetry associated with $\ga (\vec{\sig})$ (i.e. the exponentially growing part) implies that the linearized hydrodynamic field $\epsilon$, defined as ${\cal Y}^0\sim \beta \sigma^0+\epsilon$, has an exponentially growing gauge symmetry transformation which in turn implies an exponentially growing propagator $G_{\epsilon\epsilon}\propto e^{{2 \pi \ov \b} t}$. This feature leads to the exponential growth 
of finite temperature  OTOCs, $\vev{A (t) B (0) A (t) B (0)}_\b $, 
for two general scalar operators  $A,B$.  Furthermore, invariance of the system under the shift symmetry associated with $\al (\vec{\sig})$ (i.e. the  exponentially decreasing part) is crucial for the absence of the exponentially growing behavior in the 
 time-ordered correlators (TOCs), $ \vev{B (0) A (t) A (t) B (0)}_\b $. Given our identification of horizon symmetries with the symmetries of the dual quantum hydrodynamic theory, we thus see that the exponential growth of OTOCs (and the absence of the exponential growth of TOCs) for theories with a holographic dual has its gravitational origin in horizon symmetries.\footnote{Horizon symmetries in connection with quantum chaos have also been discussed in~\cite{Lin:2019qwu,Pasterski:2020xvn}.}

In \cite{Blake:2017ris,Blake:2021wqj}, the stress-energy tensor is seen as a composite operator of the hydrodynamic fields. Therefore, the presence of the exponentially growing behavior in the two-point function of $\epsilon$ may cause concerns as it may lead to the exponential growth in correlation functions of the stress tensor too. This would imply instabilities that are generically not allowed. Nevertheless, in~\cite{Blake:2017ris} it was further argued that the shift symmetries ensure that such an exponentially growing behavior does not appear in the correlation functions of the stress tensor through the phenomenon called pole-skipping~\cite{Grozdanov:2017ajz,Blake:2017ris,Blake:2018leo}. We thus conclude that the pole-skipping phenomenon in holographic theories  is also intimately related to horizon symmetries.\footnote{As we will elaborate more explicitly below, only pole-skipping in a particular channel of the stress tensor, the energy conservation channel, should be due to the horizon symmetries.}

That the horizon plays a crucial role in both the exponential growth of OTOCs and the pole-skipping phenomenon has of course been well understood. In the original calculation of OTOCs from holography~\cite{Shenker:2013pqa}, the energy of small perturbations to a thermofield double state long in the past is exponentially blueshifted from the horizon, which  
leads to the exponential growth  $e^{{2 \pi \ov \b} t}$ of the OTOCs.
The pole-skipping phenomenon was discovered on the gravity side from an analysis of the Einstein equations near the horizon in~\cite{Grozdanov:2017ajz} and elucidated in~\cite{Blake:2018leo}. 

In this section, we elaborate a bit further on these gravity calculations from the perspective of horizon symmetries.

\subsection{Connection with the shock wave geometry}
\label{sec.shockwave}
For a holographic system, an OTOC can be obtained from the dual bulk gravity system by calculating two-point functions of $B$ in a shock wave geometry generated by the operator $A$  inserted in the infinite past of a black hole geometry, see, e.g., \cite{Shenker:2013pqa,Shenker_2014multiple,Shenker:2014cwa}. We may view the shock wave as an effective description of multiple graviton exchanges between $A$ and $B$, which can in turn be interpreted in the boundary system in terms of exchanges of the hydrodynamic modes corresponding to the stress tensor.

\begin{figure}
    \centering
    \includegraphics[scale=0.7]{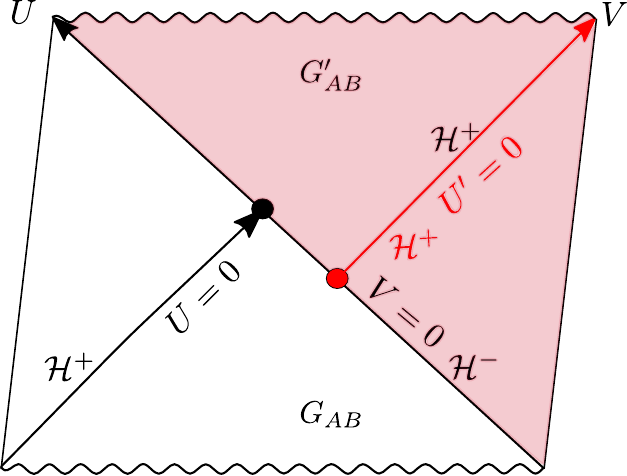}
    \caption{A geometry with a shock wave inserted at $V=0$ that leads to a discontinuity of the $U=0$ surface.}
    \label{fig:shock wave}
\end{figure}

 In this subsection, we show that the shock wave geometry of \cite{Sfetsos_1995,Shenker:2013pqa,Shenker_2014multiple,Shenker:2014cwa} (see, e.g., Fig.~\ref{fig:shock wave}) can be interpreted as being generated by a horizon symmetry transformation. 
More explicitly, we will show that near the horizon, the shock wave generated by inserting $A$ at early times may be viewed as patching together two spacetimes with metrics $G$ and $G'$ along the trajectory of 
the quantum created by $A$. Here $G$ is the black brane spacetime describing the equilibrium state, while $G'$ is related to $G$ by a horizon symmetry transformation.
In other words, we may say that an $A$ insertion generates a horizon symmetry transformation. We will now describe this connection in more detail using the example of~\eqref{fgm20}.

For this purpose it is convenient to use the  Kruzkal-Skezeres (KS) coordinates in terms of which the black brane metric~\eqref{fgm20} has the form 
\begin{equation}\label{eq:metric.UV}
	ds^2=Q (r)dUdV+g(r)d\vec{x}^2\,,
\end{equation}
where the coordinates $U$ and $V$ are given as follows
\begin{equation}
	U=-e^{-\hat \ka \left(v-2r_{*}(r)\right) }\,,\quad V=e^{\hat \ka v}\,,\quad UV=-e^{2\hat \ka r_{*}(r)}\,\quad \partial_r r_{*}(r)=F(r)^{-1}\, , 
	\label{eq:KS coord}
\end{equation}
and
\begin{equation}
	 Q(r)=\frac{F(r)}{\hat \ka^2 U V}.
	\label{eq:KS coord2}
\end{equation}
In KS coordinates, the black hole's future horizon $\cal{H}^+$ is located at $U=0$, while the past horizon $\cal{H}^-$ is at $V=0$.

The shock wave geometry $G_{\rm shock}$ generated by a high energy particle with trajectory $V=0$ similar to that used in \cite{Shenker:2013pqa,Shenker_2014multiple,Shenker:2014cwa,Sfetsos_1995}, is given by
\bega
G_{\rm shock} = G +\Theta(V){\cal L}_{\xi}G
\end{gather}
where $\Theta(V)$ is the Heaviside stepfunction. It is described by an unperturbed black brane with a metric given in equation \eqref{eq:metric.UV} for $V<0$,  while for $V\geq0$ the coordinates are shifted by $x^A\rightarrow x^A+\xi^A$ where $\xi^A$ is given by 
\bega
\label{coord.transfold}
    \xi^V=0\,,\quad 
\xi^i=0\,,\quad 
\xi^U=q\left(\vec{x}\right)\,.
\end{gather}
The function $q\left(\vx\right)$ is determined by the Einstein's equations \begin{equation}
\label{shock}
	\left(\nabla^2-d\pi T g^{\prime}(r_0)\right)q(\vec{x}')= a_0 E e^{\hat \kappa t}\delta(\vec{x}')\,.
\end{equation} 
where $a_0$ is some constant,  $E$ is the energy of the particle, and {the exponential factor $ e^{\hat \kappa t }$ comes from a boost near the horizon.

Consider now the effect of the horizon symmetry transformation~\eqref{eq:transf.Z.Y}-\eqref{eq:transf.f} parameterized by $\gamma$. Near the future horizon, this linearized\footnote{  Notice that we work at the linearized level in $\gamma(\vec{x}) V$. With this approximation, one can probe boundary times for which the backreaction due to $A$ is substantial such that we can see the exponential growth similar to \cite{Shenker:2013pqa}. This aligns with expectations from ~\cite{Blake:2017ris,Blake:2021wqj} for maximally chaotic quantum many-body systems.} transformation  can be expressed in terms of the KS coordinates as 
\bega
\label{eq:coord.transfnewV}
\chi^V={1 \ov 8 \pi } F''(r_0) \ga (\vec{x})V^2+{\cal O}(U)\,,\\ 
\label{eq:coord.transfnewx}
\chi^i ={1 \ov 2 \pi g(r_0)}  \p_i  \ga (\vec{x})V+\mathcal{O}\left(U\right)\,,\\ 
\label{eq:coord.transfnewU}
\chi^U=-\frac{ \gamma(\vec{x})}{dr_0}+\mathcal{O}\left(U\right)\,.
\end{gather}
Notice that at $V=0$, the transformation in \eqref{eq:coord.transfnewV}-\eqref{eq:coord.transfnewU} reduce to that in \eqref{coord.transfold}  upon the identification $q(\vec{x})=-\frac{ \gamma({\vec{x}})}{dr_0}$ near the future horizon.

One can use these transformations to construct a geometry with metric $G_{\rm shock}'$,
\bega
G'_{\rm shock}=G+\Theta(V)\sL_{\chi}G\,,
\end{gather}
described by the unperturbed black hole with metric \eqref{eq:metric.UV} for $V<0$, and by the metric \eqref{eq:metric.UV} with the coordinate transformation in \eqref{eq:coord.transfnewV}-\eqref{eq:coord.transfnewU} for $V\geq 0$.
Crucially, the geometry described by  $G_{\rm shock}$ and the one described by $G'_{\rm shock}$  are related by a coordinate transformation near the future horizon as follows
\bega
G_{\rm shock}=G'_{\rm shock}+\sL_{\rho}G'_{\rm shock}\,,
\end{gather}
with
\be
\label{eq:rho}
\rho =-\Theta(V)\left(\chi-\xi\right)
\ee
and  $q(\vec{x})=-\frac{ \gamma({\vec{x}})}{dr_0}$
upon linearization.
In deriving \eqref{eq:rho} we have used the fact that  the delta functions arising from derivatives on $\Theta (V)$ do not contribute as they are always multiplied by factors of $V$ near the future horizon.

Hence, near the future horizon, the two seemingly different shock wave metrics $G_{\rm shock}$ and $G_{\rm shock}'$ are related by a diffeomorphism and are thus physically equivalent. Notice that to extend this argument beyond the near-horizon limit, one needs to know how the horizon symmetry transformation~\eqref{eq:transf.Z.Y}-\eqref{eq:transf.f} acts away from the horizon. An interesting set of transformations are those for which $\rho^A\big|_{V=0}=0$. In this case, $G_{\rm shock}$ and $G'_{\rm shock}$ are related by the coordinate transformation~\eqref{eq:rho} even away from the future horizon. Consequently,  gluing of $G$ to $G+\Theta(V)\sL_{\chi}G$ using the horizon symmetry $\gamma$ serves as a solution-generating technique for the shock wave geometry $G_{\rm shock}'$ created by a highly energetic particle released from the right boundary long in the past.

The procedure outlined above can also be used to construct the shock wave geometry created by a particle released from the left boundary that follows a null trajectory close to the future horizon located at $U=0$. This geometry corresponds to gluing two halves of the unperturbed black hole along the $U = 0$ slice with a shift in the $V$ direction $V\to V+q\left(\vec{x}\right)$. To generate this kind of geometry using horizon symmetries, it is necessary to consider the symmetries of the past horizon using outgoing EF coordinates and perform the gluing similar to the construction above.

Finally, we want to comment on the special case of $d=3$. In this case, the coordinates $v$ and $V$ will not transform due to the vanishing of $F''(r_0)$ as can be seen from equation \eqref{eq:Fpp}. Nevertheless, the coordinate $U$ will transform, and thus a shockwave geometry can be constructed.  Therefore, we do expect that also for $d=3$ the OTOCs will grow exponentially due to the symmetry transformation related to $\gamma$. 

\subsection{Connections with pole skipping} 
\label{sec.poleskipping}

We now elaborate on the connection between the horizon symmetries and the phenomenon of pole skipping. 

First, we review the argument of pole-skipping contained in~\cite{Blake:2017ris}. There it was shown that the exponentially growing part of the symmetry in \eqref{hies2} parameterized by $\gamma(\vec{\sigma})$ implies  that the momentum space retarded two-point function $G_{\epsilon \epsilon} (\om, \vk)$  of the linearized hydrodynamic mode
$\epsilon$ has  a pole in the upper half complex $\omega$-plane at 
\be \label{spol}
\om = i {2\pi \ov \b}\,,
 \quad \vk^2 = - k_c^2 , \quad k_c = {2 \pi \ov \b} {1 \ov v_B} \,,
\ee
which gives rise to the coordinate space behavior
\be\label{Gen}
G_{\epsilon\epsilon}(t,\vec{x})\sim e^{\frac{2\pi}{\beta}\left(t-\frac{|\vec{x}|}{v_B}\right)}\,,
\ee
where $\frac{2\pi }{\beta}$ is the maximal Lyapunov exponent and $v_B$ is the butterfly velocity.

The stress tensor $T^{\mu \nu}$ in the hydrodynamic effective field theory approach  is expressed as a composite operator in terms of the hydrodynamic fields $\epsilon^{\mu}$ where ${\cal Y}^{\mu}\sim x^{\mu}+\epsilon^{\mu}$. Therefore its correlation functions are determined by those of $\ep^{\mu}$. However, the behavior~\eqref{Gen}, and thus the pole~\eqref{spol}, cannot appear in the two-point functions of $T^{\mu \nu}$, and thus must be canceled through how $T^{\mu \nu}$ depends on $\ep^\mu$.  Since $T^{\mu \nu}$ is a  physical observable, its dependence on $\ep^\mu$ must be invariant under the shift symmetries {parameterized by $\alpha$ and $\gamma$ contained in }\eqref{eq:transf.f}, which are gauge symmetries.  Working in the regime of no momentum conservation, it has been argued in~\cite{Blake:2017ris} that the gauge invariance of $T^{\mu \nu}$ in fact warrants that the pole~\eqref{spol} is always canceled in the retarded correlator of the energy density operator $T^{00}$ by the appearance of a zero at the same location. 
This is the phenomenon of pole-skipping.

Notice that while the pole in  $\om$ given in~\eqref{spol} is universal, the values of $k_c$ and the butterfly velocity $v_B$ depend on the specific theory and cannot be determined from symmetries. On the gravity side, their values depend on the specific background via the bulk equations of motion, see, e.g., \cite{Grozdanov:2017ajz,Blake:2018leo}. 
This can also be seen from the discussion on the shock wave geometries of the previous subsection. While the horizon symmetries allow any choice of $\ga ( \vec{\sigma})$, the explicit profile $\gamma (\vec{\sigma})$ is determined through the shock wave equation~\eqref{shock} which in turn comes from the specific form of the Einstein equations. In fact, the solution to the shock wave equation determines the spatial dependence in~\eqref{Gen}, and thus the location of the skipped pole in~\eqref{spol}.
Indeed it was shown explicitly in~\cite{Blake:2018leo} from an analysis of the Einstein equations near the horizon that precisely the shock wave equation~\eqref{shock} determines the location of $k_c$ in the pole-skipping phenomenon.
Thus horizon symmetries lead to the phenomenon of pole-skipping and specify the frequency in~\eqref{spol}, but more information is needed beyond horizon symmetries to determine $k_c$ and thus the full location of the  pole. 

We emphasize that in \cite{Blake:2017ris} the authors worked within the energy density conservation channel were the momentum dependence was explicitly turned off, that is $\epsilon^i=0$. Thus, the Green's function for $T^{00}$ was solely given in terms of its dependence on $\epsilon$ and there was no symmetry associated to $\sigma^i$ coordinates in \eqref{hies2} and \eqref{hies4}. Our prediction arising from horizon symmetries 
given in \eqref{velo2}
could play a role in the construction of a complete effective field theory for maximally chaotic systems. In particular, including momentum dependence would lead to a contribution of the propagator $G_{\epsilon^i\epsilon^i}$ even to $T^{00}$ and thus needs to be considered.  The constraints on $G_{\epsilon^i\epsilon^i}$ arising from \eqref{velo2} might be crucial  to understand the pole-skipping phenomenon in a full effective field theory for maximally chaotic systems.

 This expectation seems even more true in light of the pole-skipping phenomenon for holographic systems obtained directly from gravity calculations in \cite{Grozdanov:2017ajz,Blake:2018leo}.
There, it was shown from a near horizon analysis of the Einstein's equations of motion, that the pole-skipping phenomenon arises for every dimensionality of the bulk spacetime. Nevertheless, from our horizon symmetry analysis and from the arguments outlined in \cite{Blake:2017ris}, it would seem that in $d=3$ dimensions there would be no pole-skipping phenomenon due to the absence of an exponentially growing symmetry transformation of $\epsilon$ as can be seen in \eqref{eq:transf.f} with $F$ given by \eqref{eq:Fpp}. Thus, the prediction given in \eqref{velo2} could be pivotal in proving the pole-skipping phenomenon for holographic systems from an effective field theory approach. We reserve an explicit analysis of this point for future work.

Since the initial discussions of~\cite{Grozdanov:2017ajz,Blake:2018leo} of the pole-skipping phenomenon in the sound channel  (or energy diffusion channel in the absence of momentum conservation) associated to chaos, an infinite tower of pole-skipping points has been found in the lower half of the complex $\omega$ plane for all channels, and for many other types of fields, see, e.g., \cite{Blake:2019otz}. From the perspective of horizon symmetries considered in this paper, only the pole-skipping phenomenon associated with the energy density operator for~\eqref{spol} 
is implied by the horizon symmetries.
The pole-skipping phenomenon in the other channels/fields,  while clearly also having to do with the horizon dynamics, appears to be unrelated to the horizon symmetries identified in this work which
should only affect hydrodynamic modes. Thus we would conclude the pole-skipping phenomenon in other fields {and other gravitational channels} likely arises from different physics.

\section{Conclusions and discussions} 
\label{sec:conclusions}

In this work, we generalized the discussion of~\cite{Chandrasekaran:2018aop,Donnay:2015abr,Donnay:2016ejv} of horizon symmetries to a larger class and gave a boundary interpretation of the symmetries as the gauge symmetries of the dual quantum hydrodynamic theory in the context of the AdS/CFT correspondence. We worked out the symmetries in some simple examples and showed the existence of a new exponentially growing symmetry that has close connections with the behavior of OTOCs and the pole-skipping phenomenon. 

 In particular, we provided a new prediction for the gauge symmetries of any maximally chaotic effective field theory that has a gravitational dual. We  showed that horizon symmetry transformations can be used as a solution generating technique to build up shock wave geometries as those arising in the holographic computations of OTOCs \cite{Shenker:2013pqa,Shenker:2014cwa}. Finally, we showed that the pole in $\omega$ in the pole-skipping phenomenon is universal and implied by the horizon symmetries, while the argument for the existence of a  pole in $\vec{k}$ is model-dependent and necessitates additional bulk information. 

We now point out some interesting future directions. Among the most straightforward generalizations to consider are more general black holes, such as charged black holes and 
 those corresponding to far-from-equilibrium states. It would also be  useful to understand what happens 
to the horizon symmetries when including higher derivative or stringy corrections. The latter correspond to instances of non-maximal chaos which is subject of current ongoing research.

Another avenue is to consider spacetimes that are not asymptotically AdS since  much of our discussion can be generalized to the case of other signs of the cosmological constant. In fact, the hydrodynamic degrees of freedom can also be defined as maps between the horizon and some timelike boundary that is embedded in the interior of the bulk spacetime, not necessarely at asymptotic infinity. In this way, it is possible to analyse the hydrodynamic and chaotic regime (quantum hydrodynamic) of a putative dual theory that lives on that timelike boundary without knowing the precise details of that theory and the bulk asymptotics. See, e.g., \cite{Bredberg:2010ky,Bredberg:2011jq,Compere:2011dx,Eling:2011zz,Compere:2012mt,Pinzani-Fokeeva:2014cka} for the fluid/gravity correspondence of these so called Rindler fluids.

To have additional support to the findings of this work,  it would be useful to extend the effective field theory for maximally chaotic systems formulated in~\cite{Blake:2017ris,Blake:2021wqj} to also include momentum conservation and  derive the implications of the transformation of the spatial coordinates due to $\gamma(\vec{\sigma})$ in \eqref{eq:transf.Z.Y} on the OTOCs and the phenomenon of pole-skipping. In particular, this could possibly help the issue arising in $d=3$ we highlighted at the end of Section \ref{sec.poleskipping}. A more challenging task is to  have an explicit derivation of the chaos effective action formulated in~\cite{Blake:2017ris,Blake:2021wqj}  from holography using the methods developed in \cite{Crossley:2015tka,deBoer:2015ija}, perhaps also generalizing the findings of \cite{Eling:2016qvx}.

 In this paper, we have identified the gravitational dual to the gauge shift symmetries and the hydrodynamic symmetries. It would also be desirable to identify the gravitational dual and its relation to horizon symmetries of another symmetry appearing in the context of effective field theories of hydrodynamics: the one responsible for the entropy production \cite{Glorioso:2016gsa}. See, e.g., \cite{Marjieh:2021gln,Bu:2022esd} for results in this direction.  At the same time, one might wonder whether it is possible to identify gravitational and effective field theory symmetries related to the additional, infinite set of skipped poles found in \cite{Blake:2019otz} in various channels and for various fields.

It has been argued that horizon symmetries  lead to local charges upon using the Wald-Zupas \cite{Wald:1999wa} or the Barnich and Brandt formalism \cite{Barnich:2001jy} (see, e.g., \cite{Donnay:2015abr,Donnay:2016ejv,Chandrasekaran:2018aop}).  It would be interesting to explore whether our horizon symmetries lead to local charges and what are their implications.
In particular, local charges have to satisfy an integrability condition, that is they should be independent of the path taken in the so-called phase space, see, e.g., \cite{Ciambelli:2022vot}, for details.
This problem has been circumvented in  \cite{Donnelly:2016auv} by introducing edge modes:  extra dynamical fields that ensure the resulting charges always respect the integrability condition. 
It would be interesting to explore possible relations between the gravitational edge modes of \cite{Donnelly:2016auv} and the fluid dynamic degrees of freedom described in these notes.

\section*{Acknowledgments}

We would like to thank Ben Craps, Juan Hernandez, and Mikhail Khramtsov for useful discussions. 	
NPF is supported by the European
Commission through the Marie Sklodowska-Curie Action UniCHydro (grant agreement ID: 886540). NPF would also like to acknowledge support from the Center for Theoretical Physics and the Department of Physics at the Massachusetts Institute of Technology where most of the work was carried out. NPF was also supported in  part by the National Science Foundation under the Grant No. PHY-1748958 to the Kavli Institute for Theoretical Physics (KITP) during the workshop refluids23 where part of the work was done. MK is supported by FWO-Vlaanderen project G012222N and by the VUB Research Council through the Strategic Research Program High-Energy Physics. HL is supported by the Office of High Energy Physics of U.S. Department of Energy under grant Contract Number  DE-SC0012567 and DE-SC0020360 (MIT contract \# 578218).

\appendix

\section{Freedom in choosing $n^A$}
\label{A.nA}

Having defined $\hat{\ell}^a$ as the unique vector that satisfies $h_{ab}\hat{\ell}^a =0$ on the horizon ${\cal H}$ up to rescalings \eqref{eq:rescaling}, there is yet a freedom in choosing the vector $n_A$. This can be shown as follows.
Consider  a basis of tangent vectors to $\sM$  consisting  of $(\ell^A, n^A, V^A_i)$ satisfying on $\sH$  
\be 
G_{AB} \ell^A V_i^B \heq 0, \quad G_{AB} n^A V_i^B \heq 0, \quad G_{AB} V^A_i V_j^B \heq \de_{ij} \,,
\ee
as well as the usual relations
\be
 G_{AB}n^A n^B \heq 0\,,\qquad G_{AB}\ell^A n^B\heq -1\,.
\ee
We can always reparameterize $n^A$ and therefore $V_i^A$ on ${\cal H}$ at will, so long as the above conditions are left invariant.

 The most general transformation of this kind $n^A\rightarrow \tilde{n}^A$ and $V_i^A \rightarrow \tilde{V}_i^A$ can be parameterized as follows
\be 
\tilde n^A \heq  n^A + a \ell^A + c_i V^A_i, \quad \tilde V^A_i \heq d_{ij} V^A_j+  b_i \ell^A , 
\ee
then 
\bega 
\ell^A \tilde{V}_{A\,i}\heq 0\,,\quad \ell^A \tilde n_A \heq-1, \quad \tilde n^A \tilde n_A \heq - 2 a + c_i^2, \quad \tilde n_A \tilde V^A_i \heq - b_i  
+ d_{ij} c_j\,,  \\
 \tilde V^A_i \tilde V_{Aj} \heq d_{ik} d_{jk}  \ .
\end{gather} 
We thus need $d_{ij}$ to be an orthogonal matrix, which we can take to be the identity matrix. The other quantities are then set as
\be \label{repf}
 d_{ij}\heq \delta_{ij}\,,\quad a \heq \ha W^2, \quad c_i = b_i \heq W_A V^A_i ,
\quad W^A \ell_A \heq W^A n_A \heq 0 \,.
\ee

Thus, to summarize, we may always redefine $n_A$ on the horizon as follows
\bega
\label{eq:rep.nA}
 n_A \rightarrow \tilde{n}_A=n_A +\Phi_A \,,\\
\Phi^A \heq \hat{\Phi}^A \heq W^A +\frac{1}{2}\ell^AW^2\,,\quad W^A\ell_A\heq 0\,,\quad W^An_A \heq 0\,,
\end{gather}
where $\Phi^A$ is a generic vector while $\hat{\Phi}^A$ is its value evaluated at the horizon ${\cal H}$ which can be decomposed into a part that is parallel to $\ell^A$ and a part that is orthogonal to it. The vector $W^A$ has to be taken as a function of the horizon coordinates.
The redefinition \eqref{eq:rep.nA} implies a redefinition of the projector $q_A{}^B$ too,
\bega
\label{eq:rep.q}
  q_{A}{}^{B}\rightarrow \tilde{q}_{A}{}^{B}= q_A{}^{B}+\Phi_A\ell^B +\ell_A\Phi^B\,,
\end{gather}
which will be useful later on.

The reparameterization freedom \eqref{eq:rep.nA} and \eqref{eq:rep.q} is akin to the Carrollian boost transformation which arises in the context of ultra relativistic field theories. It has been shown that it is also a symmetry of null surfaces when they are embedded in a Lorentzian spacetime of one dimension higher, see, e.g., \cite{Hartong:2015xda}.

\subsection{Independence of $n^A$} 

The arbitrariness in choosing $n^A$ seems to naively  affect the horizon constraint equations derived in \eqref{Constr1}, \eqref{Constr2} as well as in \eqref{ss1}, \eqref{acd3}, and \eqref{fi1}. However, we will now show that this is not the case.

{$\bullet$ $\chi^A \ell_A \heq 0$}

We start with the case where $\chi^A$ is tangent to the horizon, that is  $\chi^A\ell_A\heq 0$.
 In this case, $\chi^A\heq \partial_a X^A \xi^a$ and $\xi^a$ is decomposed as in \eqref{decomp.chi}. Given that $\hat{n}_a\heq \partial_aX^A n_A$, the reparameterization freedom of $n_A$ obtained in \eqref{eq:rep.nA} implies a reparameterization freedom of  $\hat{n}_a$ and $\hat{q}_a{}^b$ as follows
\be
\hat{n}_a\rightarrow \tilde{\hat{n}}_a=\hat{n}_a+\hat{\Phi}_a\,,\quad
\hat{q}_a{}^b \rightarrow \tilde{\hat{q}}_a{}^b =\hat{q}_a{}^b+\hat{n}_a\hat{\ell}^b \,,
\ee
with
\be 
\hat{\Phi}_a\heq \partial_aX^A \Phi_A\,,\quad\hat{\Phi}_a\hat{\ell}^a=0\,,\quad \hat{q}_a{}^b =\delta_a{}^b+ \hat{n}_a \hat{\ell}^b\,.
\ee

Given that $\xi^a$ is a generic diffeomorphism on the horizon and should be independent of the chosen $\hat{n}_a$, this  arbitrariness affects the parameterization \eqref{decomp.chi} via the following redefinitions
\be
\tilde{f} = - \xi^a \tilde{\hat{n}}_a = f - h, \quad Y^a \to \tilde{Y}^a =\xi^b \tilde{\hat{q}}_b{}^a = Y^a + h \hat \ell^a, \quad a
\to \tilde{a} = a + \sL_{\hat \ell} h \,,
\ee
with
\be h = Y^a  \hat{\Phi}_a \ .
\ee
It can be readily checked that equation~\eqref{dvq} is equivalent to that with $f, a, Y^a$ replaced by tilded quantities. In this way, it is clear that equations \eqref{Constr1} and \eqref{Constr2} (as well as \eqref{dvq}) do not depend on the specific choice of $\hat{n}_a$ as expected given the  equivalence to~\eqref{eq:lb}--\eqref{eq:kb}  that  are explicitly independent of it.

{$\bullet$ $\chi^A\ell_A \neq 0$ }

Consider now the case with $\chi^A\ell_A \neq 0$. The most general parameterization of $\chi^A$ is given in \eqref{Znonzero}, and under the redefinition of $n_A$ and $q_A{}^B$ given in \eqref{eq:rep.nA} and \eqref{eq:rep.q}, we have 
\bega
\label{eq:red.f}
f\rightarrow \tilde{f}\heq -\chi^A \tilde{n}_A\heq f-H\,,\\
\label{eq.red.Y}
Y^A\rightarrow \tilde{Y}^A\heq \tilde{q}^A{}_B\chi^B \heq Y^A+H\ell^A-Z\Phi^A\,,\\
\label{eq:red.Z} 
Z\rightarrow \tilde{Z}\heq -\chi^A \ell_A\heq Z \,,
\end{gather}
with 
\bega
\Phi_A\heq \hat{\Phi}_A \heq W_A+\frac{1}{2}W^2\ell_A\,,\quad W^A\ell_A \heq 0\,,\quad W^An_A\heq 0\,,\\
 H=h-\frac{1}{2}W^2Z\,,\quad h=Y^AW_A\,.
\end{gather}
The  following  expressions will be useful
\begin{equation}
\Phi_A\ell^A\heq 0 \,,\quad \Phi_An^A \heq  -\frac{1}{2}W^2\,,\quad \Phi^2\heq W^2\,.
\end{equation}

The symmetry constraint equation \eqref{ss1} is clearly invariant under \eqref{eq:rep.nA} and \eqref{eq:red.Z}. 
 We now want to show that the expression \eqref{acd3} is also covariant under the above transformations, that is it has the same form once all the quantities have been replaced by tilded ones. To do so, let us perform the intermediate transformations evaluated at the horizon
 \bega
 \label{eq:transf1}
 \sL_{\ell}Y^A \rightarrow \sL_{\ell}Y^A+\ell^A\sL_{\ell}H-\kappa Z \hat{\Phi}^A -Z\sL_{\ell}\hat{\Phi}^A\,,\\
  -q^{AB}\nabla_BZ\rightarrow  -q^{AB}\nabla_BZ -\kappa Z\hat{\Phi}^A -\ell^A \sL_{\hat{\Phi}}Z\,,\\
 Z\eta^A_{\perp}\rightarrow  Z\eta^A_{\perp}+2\kappa Z \hat{\Phi}^A +Z\ell^A \eta^B_{\perp}W_B
+\kappa ZW^2 \ell^A
+Z\sL_{\ell}\hat{\Phi}^A+Zn_B\ell^A\sL_{\ell}\hat{\Phi}^B+Z\ell^A\hat{\Phi}_B\sL_{\ell}\hat{\Phi}^B\,,\\
\label{eq:transf4}
a\rightarrow a+\sL_{\ell}H+Zn_A\sL_{\ell}\hat{\Phi}^A-\sL_{\hat{\Phi}}Z
+Z\eta^A_{\perp}W_A +\kappa Z W^2 +Z\hat{\Phi}_A \sL_{\ell}\hat{\Phi}^A\,,
 \end{gather}
where we used \eqref{ss1} and \eqref{acd3} to simplify the expressions, the fact that $\ell^A \nabla_A$ is a derivative along the horizon, as well as $\eta^A \ell_A \heq 2\kappa$,   $\eta^A \Phi_A \heq \eta^A_{\perp}W_A +\kappa W^2$, $\sL_{\ell}\Phi^A \heq \sL_{\ell}\hat{\Phi}^A$, and
\be 
\label{eq:some.rel}
q^{AB}\sL_{\Phi}\ell_A\heq 0\,,\quad \ell^A\sL_{\Phi}\ell_A\heq 0\,,\quad \Phi^A\sL_{\Phi}\ell_A\heq 0\,,\quad \ell_A\sL_{\Phi}\ell^A\heq 0\,.
\ee
These latter equations can be easily derived using the fact that $\ell^2\heq 0$, $q^{AB}\nabla_B$ and $\ell^B\nabla_B$ are derivatives along the horizon, and the fact that $\ell_A$ is hypersurface orthogonal, that is 
\be
q^{AB}\sL_{\Phi}\ell_A = q^{AB}(\Phi^C\nabla_C \ell_A +\ell_C\nabla_A \Phi^C)\heq q^{AB}W^C(\nabla_C \ell_A -\nabla_A \ell_C)\heq 0\,.
\ee
Combining the transformations \eqref{eq:transf1}-\eqref{eq:transf4} we find that \eqref{acd3} is equivalent to the same equation with tilded variables.

Let us now consider \eqref{fi1} which we copy here for convenience
\bega
\sL_{\ell}(\sL_{\ell}+\kappa)f+(\sL_{\ell}+\kappa)a+\sL_{Y}\kappa+\eta^{\perp \, A}\nabla_AZ +Z{\cal F}\heq 0\nonumber\,,\\
{\cal F}=\frac{1}{2}\ell^A \ell^B\sL_n\sL_nG_{AB} -\eta^2-\sL_{\ell}\lambda-3\kappa\lambda\,.
\end{gather}
and we have performed a slight rewriting of ${\cal F}$ using 
\begin{equation}
\eta^2_{\perp}\heq \eta^2 -4\kappa \lambda\,.
\end{equation} 
Under the redefinition \eqref{eq:rep.nA},
the various terms, when evaluated on the horizon ${\cal H}$, transform as follows
\bega
\sL_{\ell}(\sL_{\ell}+\kappa)f\rightarrow\sL_{\ell}(\sL_{\ell}+\kappa)f -\sL_{\ell}(\sL_{\ell}+\kappa)H\,,\\
(\sL_{\ell}+\kappa)a \rightarrow (\sL_{\ell}+\kappa)a+ \sL_{\ell}\sL_{\ell}H+\kappa \sL_{\ell}H+2\kappa Zn_A\sL_{\ell}{\hat{\Phi}}^A+Z\sL_{\ell}( n_A \sL_{\ell}\hat{\Phi}^A)+\nonumber\\
 -Z\sL_{\hat{\Phi}}\kappa- 2\kappa \sL_{\hat{\Phi}}Z+2\kappa Z \eta^{\perp A}W_A+Z\sL_{\ell}(\eta^{\perp A}W_A)+2\kappa Z \hat{\Phi}_A\sL_{\ell}\hat{\Phi}^A+Z\sL_{\ell}(\hat{\Phi}_A \sL_{\ell}\hat{\Phi}^A)+\nonumber\\
 -\sL_{\ell}\hat{\Phi}^A \nabla_AZ+Z\sL_{\ell}(\kappa W^2)+2\kappa^2 W^2  Z\,,
\\
\sL_{Y}\kappa \rightarrow \sL_Y\kappa +H\sL_{\ell}\kappa-Z\sL_{\hat{\Phi}}\kappa\,,\\
\eta^A\rightarrow \eta^A+G^{AB}\sL_{\Phi}\ell_B-\sL_{\Phi}\ell^A\,,\\
\eta^{\perp\, A}\nabla_A Z\rightarrow \eta^{\perp\, A}\nabla_AZ+\kappa Z \eta^{\perp\, B}W_B+2\kappa \sL_{\hat{\Phi}}Z+\kappa^2 W^2Z+\sL_{\ell}\hat{\Phi}^B  \nabla_BZ+\nonumber\\
+\kappa Zn_B\sL_{\ell}{\hat{\Phi}}^B +\kappa Z\hat{\Phi}_B\sL_{\ell}\hat{\Phi}^B \,,
\\
\frac{1}{2}\ell^A\ell^B \sL_{n}\sL_n G_{AB}\rightarrow  \frac{1}{2}\ell^A\ell^B \sL_{n}\sL_n G_{AB} +2\sL_{\hat{\Phi}}\kappa -2\eta_A\sL_{\hat{\Phi}}\ell^A -\delta \eta_A \sL_{\hat{\Phi}}\ell^A-\sL_{\ell}{\cal A} +\kappa {\cal A}\,,\\
\eta^2 \rightarrow \eta^2 +4\kappa {\cal A} -2\eta_A\sL_{\hat{\Phi}}\ell^A-\delta \eta_A \sL_{\hat{\Phi}}\ell^A\,,\\
\lambda \rightarrow \lambda -{\cal A}+n_A\sL_{\ell}\hat{\Phi}^A+\eta^A_{\perp}W_A +\kappa W^2+\hat{\Phi}_A \sL_{\ell}\hat{\Phi}^A\,,\\
\sL_{\ell}\lambda\rightarrow \sL_{\ell}\lambda -\sL_{\ell}{\cal A}+\sL_{\ell}(n_A\sL_{\ell}\hat{\Phi}^A)+\sL_{\ell}(\eta^A_{\perp}W_A) +\sL_{\ell}(\kappa W^2)+\sL_{\ell}(\hat{\Phi}_A \sL_{\ell}\hat{\Phi}^A)\,,
\end{gather}
where we used 
\bega
\sL_{n}\sL_{\Phi}G_{AB} = \sL_{\Phi}\sL_{n}G_{AB}+\sL_{[n,\Phi]}G_{AB}=\sL_{\Phi}\sL_{n}G_{AB}+2\nabla_A(G_{BC}\sL_{n}\Phi^C )\,,\\
\ell_C\sL_n\Phi^C = -\ell_C\sL_{\Phi}n^C\heq n^C\sL_{\Phi}\ell_C \heq -{\cal A}\,,
\end{gather}
and we have defined
\be
\sL_{\Phi}\ell_C \heq {\cal A}\ell_C \,,\quad {\cal A} \heq - n^C\sL_{\Phi}\ell_C \,,
\ee
using \eqref{eq:some.rel}. In all the steps above we repeadetely used $\ell^A n_A \heq -1$, $\ell^2 \heq 0$, $\ell^A \hat{\Phi}_A \heq 0$, $\delta \eta_A=\sL_{\Phi}\ell_A-G_{AB}\sL_{\Phi}\ell^B $, $\eta^A \ell_A \heq 2\kappa$, $\eta^A n_A \heq \lambda$, $\delta \eta^B \ell_B \heq 0$, and   the fact that various derivatives are taken to be along the horizon.

Combining all the results above, it is straightforward to show that equation \eqref{fi1} is equivalent to the one with the tilded variables as we wanted to demonstrate.

\section{Details of some derivations} \label{app:der}

{\bf $\bullet$ Proof of~\eqref{ej1}:}

To derive equation~\eqref{ej1}, first  we note that 
\bega
\ell^B \de \nab_B \ell^A  = 
\ell^B \de \Ga^A_{BC} \ell^C  = \frac{1}{2} \ell^B\ell^C (\nabla_B f^A_{C}+\nabla_{C}f^A_{B}-\nabla^A f_{BC}) \nonumber \\
=  \ell^B\ell^C \nabla_B f^A_{C}-\ha  \ell^B\ell^C  \nabla^A f_{BC}\,.
\label{y00} 
\end{gather} 
Assuming~\eqref{1''}, we have 
\bega \label{y10}
\ell^B\ell^C \nabla_B f^A_{C} =\ell^B \nabla_B (f^A_{C} \ell^C) - \ell^B  f^A_{C} \nabla_B  \ell^C
\heq - \ell^A \sL_\ell c,  \\
\ell^B\ell^C \nabla_A  f_{BC}
\heq \nabla_A ( f_{BC} \ell^B \ell^C) 
+ 2 c  \,  \ell_C  \nabla_A  \ell^C  \heq  \p_A ( f_{BC} \ell^B \ell^C + c \, \ell_C    \ell^C ) \ . 
\label{y20}
\end{gather} 
where we used the fact that $\ell^2 =\ell_C\ell^C \heq 0$.
Contracting these expressions with $\ell_A$ or $q_{A}{}^D$ leads to a vanishing result given that $f_{BC}\ell^B\ell^C\heq 0$ (due to \eqref{1''}), $\ell^2\heq 0$, and derivatives along $\ell^A \nabla_A$ and $q^{AD}\nabla_A$ are along the horizon surface and therefore vanish if their argument is vanishing along the horizon.

 Contracting \eqref{y10} and \eqref{y20} along $n_A$ instead leads to a nontrivial result. We get
\bega 
n_A\ell^B\ell^C\nabla_Bf^A_C \heq \sL_{\ell}c\,,\\
n^A \ell^B\ell^C \nabla_A  f_{BC} \heq \sL_n ( f_{BC} \ell^B \ell^C + c \, \ell_C    \ell^C ) \cr
=  \ell^B \ell^C \sL_n f_{BC} - c \ell_A \sL_n \ell^A + c \ell^A \sL_n \ell_A
=  \ell^B \ell^C \sL_n f_{BC} + 2 \ka c
\label{y30}
\end{gather}
where we have used that 
\be
 \ell^A \sL_n \ell_A -  \ell_A \sL_n \ell^A = 2 \ell^A \ell_B \nab_A n^B \heq  2 \ka \ .
 \ee
as well as $n_A\ell^A\heq -1$ and the fact that $\ell^A\nabla_A$ is a derivative along the horizon hypersurface. Combining these two expressions into \eqref{y00} we get \eqref{ej1} as desired.

{\bf $\bullet$ Proof of~\eqref{ss1}, \eqref{acd3}, \eqref{fi1}, and \eqref{fi2}:}

Here we give details of the derivations of~\eqref{ss1}--\eqref{acd3}, \eqref{fi1}, and~\eqref{fi2}.  
For this purpose, we copy equations~\eqref{cd1}--\eqref{cd10} and~\eqref{ej1}  with the decomposition \eqref{ej3} here for convenience.
\bega\label{acd1}
 \ell^A  \sL_\chi G_{AB}=  \sL_\chi \ell_B -G_{AB}  \sL_\chi \ell^A  
 = \sL_\chi \ell_B + G_{AB}  \sL_\ell \chi^A
 \propto \ell_B  \, , \\
 \label{aej1}
\frac{1}{2}\sL_{\chi}(\ell^A \ell^B\sL_{n}G_{AB})- \ell^B (\sL_{\chi}\ell^A)  (\sL_nG_{AB})+ \frac{1}{2}\ell^A \ell^B\sL_{[n,\chi]}G_{AB}+ c \ka  - \sL_\ell c = 0 \,, \\
\label{acd10} 
c \equiv \ell^A n^B \sL_\chi G_{AB} = n^B \sL_\chi \ell_B - n_A \sL_\chi \ell^A   \ .
\end{gather}

Using the decomposition of the diffeomorphisms vector $\chi^A$ given in  \eqref{Znonzero}, we can separate the equations~\eqref{acd1}--\eqref{acd10} into the parts involving $\tilde \chi^A$ and the parts involving $Z n^A$. 
Consider first~\eqref{acd1}. It can be written as 
\bega \label{acd2intermit}
 Z \sL_{n} \ell_B  - \nab_B Z 
  +G_{AB}  \sL_\ell Y^A -  Z G_{AB} \sL_n \ell ^A 
+ n_B \sL_\ell Z  + \sL_{\tilde \chi} \ell_B
 \propto  
 \ell_B   \ ,
 \end{gather}
and  further simplified as 
 \bega \label{acd2}
 Z \eta_B 
 + 2 n_B \sL_\ell Z  - q_B^A \nab_A Z  +G_{AB}  \sL_\ell Y^A   
  \propto \ell_B \,,
 \end{gather}
 where we have used the definition \eqref{ceh2} and $\sL_{\tilde \chi} \ell_B \propto \ell_B$ as discussed in~\eqref{cd21}. 
 
 Multiplying both sides of~\eqref{acd2} by $\ell^B$ and using 
 \bega 
 \ell^A \eta_A = \ell^A \ell^B  (\sL_nG_{AB})  = 2  \ell^A \ell^B \nab_A n_B \heq - 2\ell^A n_B \nab_A \ell^B \heq 2 \ka \\
\ell_A  \sL_\ell Y^A = - Y^A \sL_\ell \ell_A  \heq 0\,, 
 \end{gather} 
 where again we used $Y^A\ell_A\heq 0$ and the fact that $\ell^A \nabla_A$ is a derivative along the horizon,
  we find that 
\be \label{sss1} 
\sL_{\ell}Z-\ka Z\heq 
  0  \ ,
 \ee
 which is \eqref{ss1}.
 
Using~\eqref{sss1} in~\eqref{acd2} then gives
\bega  \label{sacd3}
 \sL_\ell Y^A  -  q^{AB} \nab_B Z + Z\eta^A + 2 \ka Z n^A \heq  \propto \ell^A  ,
 \end{gather}
 which, using~\eqref{ceh2} and the fact that $\eta^A\ell_A \heq 2\kappa$, can be further simplified to 
 \be
  \sL_\ell Y^A  -  q^{AB} \nab_B Z + Z\eta^A_\perp  \heq  
 a   \ell^A \,,
 \ee
 which is the desired equation \eqref{acd3},
where $a$ can be written as  
 \be \label{ceh}
 a =  - n_A \sL_\ell Y^A   \ .
 \ee
 
 Finally, to prove  \eqref{fi1},
  consider~\eqref{aej1}, whose LHS can be written as 
 \be 
 {\rm LHS} = {\rm LHS}_1 + {\rm LHS}_2\,,
 \ee
 where $ {\rm LHS}_1$ denotes the part involving $\tilde \chi^A$ and ${\rm LHS}_2$ denotes the part involving $Z n^A$. 
First note that 
\bega 
c = n^B \sL_{\tilde \chi} \ell_B + n^B \sL_{Zn } \ell_B + n_A \sL_\ell \tilde \chi^A + n_A \sL_\ell (Zn^A) \cr
\heq \tilde c + Z (\sL_n (n^B  \ell_B) + n_A  \sL_\ell n^A ) -  \sL_n Z    
\heq \tilde c + Z \lam - \sL_n Z , \\
\tilde c \equiv n^B \sL_{\tilde \chi} \ell_B  + n_A \sL_\ell \tilde \chi^A ,   \\
-\sL_\ell c \heq  -\sL_\ell \tilde c - \sL_\ell (Z \lam  -  \sL_n Z  ) 
\heq -\sL_\ell \tilde c  - Z ( \sL_\ell  + \ka ) \lam + \sL_\ell \sL_n Z  \,,
\end{gather}
where we used the fact that $n^2=n_An^A \heq 0$, that $\ell^A \nabla_A$ is a derivative along the horizon, and equation \eqref{sss1}.
The parts of the terms in~\eqref{aej1} that involve $Z n^A$ have the form 
\bega
\ha  \sL_{\chi}(\ell^A \ell^B\sL_{n}G_{AB})  = Z n^C \nab_C (\ell^A \ell^B \nab_A n_B )  + \cdots 
= \ha Z \sL_n (\ell^A \eta_A) + \cdots \,,\\
 -\ell^B (\sL_{\chi}\ell^A)  (\sL_nG_{AB}) \heq +Z \eta_A (\sL_{\ell} n^A + \ka n^A)  + \cdots 
  = + Z \eta_A \sL_{\ell} n^A  +  \ka Z \lam + \cdots
 \,, \\
 \ha  \ell^A \ell^B\sL_{[n,\chi]}G_{AB} = \ell^A \ell^B \nab_A [n, Z n]_B + \cdots \,,\\
 = \ell^A \ell^B  \nab_A (n_B \sL_n  Z) +\cdots 
 \heq - \sL_\ell \sL_n Z + \ka  \sL_n  Z +\cdots \,,
\end{gather} 
where $\cdots$ denote the parts involving $\tilde \chi^A$ and we used repeadetely $\ell^An_A \heq -1$, $n^2\heq 0$, and the fact that $\ell^A \nabla_A$ is a derivative along the horizon. From the above equations we then find 
\bega
{\rm LHS}_2 \heq \ha Z \sL_n (\ell^A \eta_A)
+ Z \eta_A \sL_{\ell} n^A
+ \ka Z \lam  - \sL_\ell \sL_n Z + \ka  \sL_n  Z  
 \cr
+ \ka (Z \lam - \sL_n Z) 
-  Z ( \sL_\ell  + \ka ) \lam + \sL_\ell \sL_n Z
\cr
= \ha Z \sL_n (\ell^A \eta_A)
+Z \eta_A \sL_{\ell} n^A 
-  Z  (\sL_\ell -\ka)   \lam \cr
= \ha Z \ell^A \ell^B \sL_n \sL_n G_{AB}  -  Z  (\sL_\ell -\ka)   \lam
 \ .
\label{lhs2}
\end{gather}

For the $\tilde \chi$ part of~\eqref{aej1}, we have 
\bega \label{acc1}
 {\rm LHS}_1 = \frac{1}{2}\sL_{\tilde{\chi}}(\ell^A \ell^B\sL_{n}G_{AB})- \ell^B (\sL_{\tilde{\chi}}\ell^A)  (\sL_nG_{AB})+ \frac{1}{2}\ell^A \ell^B\sL_{[n,\tilde{\chi}]}G_{AB}+ \tilde{c} \ka  - \sL_\ell \tilde{c}\,,
 \end{gather}
 where
 \bega
 \tilde c \equiv n^B \sL_{\tilde \chi} \ell_B  + n_A \sL_\ell \tilde \chi^A = -\tilde b + n_A \sL_\ell \tilde \chi^A , \\
\sL_{\tilde \chi} \ell_B \heq \tilde b \ell_B , \quad
   \tilde b \heq - n^A \sL_{\tilde \chi} \ell_A = \ell_A \sL_{\tilde\chi} n^A  = - \ell_A \sL_n \tilde \chi^A  \ .
\end{gather} 
From~\eqref{acd3} we get
\bega \label{ne12}
\sL_\ell \tilde \chi^A  \heq \ell^A \sL_\ell f + \sL_\ell Y^A  \heq (\sL_\ell f +  a) \ell^A +  q^{AB} \nab_B Z - Z  \eta^A_\perp 
\\
n_A \sL_\ell \tilde \chi^A  =  - (\sL_\ell f +  a )\,, \quad  
\Lra \quad \tilde c = - (\tilde b + \sL_\ell f +  a)  
\ .
\end{gather}
Now note that 
\bega
 \ha \sL_{\tilde \chi}(\ell^A \ell^B\sL_{n}G_{AB})
\heq  \sL_{\tilde \chi}\kappa\,, \\ 
\ha \ell^A \ell^B\sL_{[n,\tilde \chi]}G_{AB} =  \ell_A\ell^B \nabla_B \sL_{n}\tilde \chi^A \heq
-  \sL_\ell  \tilde b -  \sL_{n} \tilde \chi^A \ell^B \nabla_B \ell_A 
\heq -  \sL_\ell \tilde b  +  \tilde b\ka \,,
\end{gather} 
and from~\eqref{ne12} we have 
\bega 
 -\ell^B (\sL_{\tilde \chi}\ell^A)  (\sL_nG_{AB}) =  +  \eta_A \sL_{\ell} \tilde \chi^A   
 \heq + \eta_A ((\sL_\ell f + a) \ell^A +  q^{AB} \nab_B Z - Z  \eta^A_\perp )
  \cr
 \heq + 2 (\sL_\ell f + a) \ka + \eta^A_\perp \nab_A Z   - Z \eta^2_\perp , \quad   
 \end{gather} 
 In deriving these steps we used several times the fact that $n_A\ell^A \heq -1$ and that $\tilde{\chi}^A \nabla_A$ is a derivative along the horizon.
We thus find that equation~\eqref{acc1} becomes  
\bega 
 {\rm LHS}_1 = \sL_\ell ( \sL_\ell + \ka)  f  + ( \sL_\ell + \ka) a + 
\sL_{Y}\kappa + \eta^A_\perp \nab_A Z   - Z \eta^2_\perp 
  \ .
 \end{gather} 
 Combining with~\eqref{lhs2} we find 
   \bega \label{fi11}
 \sL_\ell ( \sL_\ell + \ka)  f  + ( \sL_\ell + \ka)  a + 
\sL_{Y}\kappa + \eta_A^\perp \nab_C Z  + Z {\cal F}  = 0 , \\
{\cal F} =  \ha  \ell^A \ell^B \sL_n \sL_n G_{AB} - \eta^2_\perp  - (\sL_\ell -\ka)  \lam 
 \,
  \end{gather}  
 which is the desired equation \eqref{fi1}.

Alternatively, using~\eqref{ceh} we can write~\eqref{fi11} as    
  \bega \label{fi21}
 \sL_\ell ( \sL_\ell + \ka)  f    + 
 Y^A ( \sL_\ell + \ka)   \sL_\ell n_A +
\sL_{Y}\kappa + q^{AC}  \nab_C Z  (\eta_{\perp A} + \sL_\ell n_A) + Z{\Tilde{\cal F}}  =0 ,\\
{\Tilde{\cal F}} ={{\cal F}}    -  \eta_\perp^A    \sL_\ell n_A  
=  \ha  \ell^A \ell^B \sL_n \sL_n G_{AB}   - \eta^A_\perp (\eta_{\perp A} + \sL_\ell n_A)
- (\sL_\ell -\ka)  \lam  \ ,
 \label{fi31}
  \end{gather} 
  where we used \eqref{acd3}, the identity $\ell^A\sL_{\ell}n_A\heq 0$, the fact that $Y^An_A\heq0$, and that $\ell^A \nabla_A$ is a derivative along the horizon.
  Note that 
  \bega 
  q^{AC} \le(\eta_{\perp C} + \sL_\ell n_C  \ri)  = q^{AB} \ell^C(\nab_B n_C +\nab_C n_B) +q^{AC} ( \ell^B \nab_B n_C + n_B \nab_C \ell^B) 
  \cr
= 2 q^{AB} \ell^C \nab_C n_B \ .
  \end{gather} 
  So, equations~\eqref{fi21}--\eqref{fi31} can be further simplified to 
 \bega \label{fi22}
 \sL_\ell ( \sL_\ell + \ka)  f    + 
 Y^A ( \sL_\ell + \ka)   \sL_\ell n_A +
\sL_{Y}\kappa + 2 q^{AB}  \nab_A Z  \ell^C \nab_C n_B+ Z {\Tilde{\cal F}} =0 ,\\
{\Tilde{\cal F}} =  \ha  \ell^A \ell^B \sL_n \sL_n G_{AB}   + 2   n_A \ell^C \nab_C  \eta^A_\perp 
- (\sL_\ell -\ka)  \lam  \ ,
 \label{fi32}
  \end{gather} 
  which is \eqref{fi2}, where again we used $\eta^{A}_{\perp}n_A \heq 0$ and the fact that $\ell^A \nabla_A$ is a derivative along the horizon.

\section{Different descriptions of diffeomorphism transformations}  \label{app:A}

Consider a submanifold $N$ embedded in $M$ described by $X^A (\xi^a)$ where $x^A$ denote coordinates of $M$ and $\xi^a$ denote coordinates of $N$. 
Consider 
\be 
G_{AB} \to G_{AB}' = G_{AB} + \sL_\chi G_{AB} 
\ee
which gives 
\be 
\label{habp}
h'_{ab} = h_{ab} +  (\nab_A \chi_B + \nab_B \chi_A)  \p_a X^A \p_b X^B \ .
\ee
Alternatively, we can take $G_{AB}$ to be fixed while changing the embedding $X^A \to X'^A = X^A +\chi^A $
under which 
\bega 
h'_{ab} 
= h_{ab} + G_{AB} (\p_a \chi^A \p_b X^B + \p_a X^A \p_b \chi^B )+ \p_C G_{AB} \chi^C  \p_a X'^A \p_b X'^B\,,
\end{gather}
which is equivalent to \eqref{habp}.

Now consider a deformation of $N$ by taking $X^A (\xi) \to X'^A = X^A + \de X^A$. The restriction $\phi (\xi)$ of a scalar function $\Phi (x)$ to $N$ now changes to 
\be 
\phi' (\xi) = \Phi (X'^A (\xi)) \,,
\ee
and the induced metric on $N$ now has the form 
\be 
h'_{ab} d\xi^a d \xi^b = G_{AB} (X') \p_a X'^A \p_b X'^B \ .
\ee
Alternatively, we can keep the embedding $X^A$ fixed and consider a diffeomorphism transformation on $\Phi$ and $G_{AB}$
\bega 
\phi' (\xi) = \Phi' (X (\xi)) , \quad h'_{ab} d\xi^a d \xi^b = G'_{\mu \nu} (X) \p_a X^A \p_b X^B, \\
 \Phi' (x) = \Phi (x') ,  \quad
G'_{AB} (x)  = G_{CD}(x') \p_A x'^C \p_B x'^D  \ .
\end{gather} 
The two descriptions are completely equivalent. 

\bibliography{HorizonSymmetries}

\end{document}